%% file: main.tex
\documentclass[ALICE,manyauthors]{cernphprep}
\usepackage[comma,square,numbers,sort&compress]{natbib}
\usepackage{hyperref}
\usepackage{upgreek}
\usepackage{lineno}
\usepackage{xspace}
\usepackage{tikz}
\usepackage[T1]{fontenc}
\usepackage{orcidlink}

\begin{document}
\input{commands.tex}

\begin{titlepage}
\PHyear{2023}       
\PHnumber{267}      
\PHdate{20 November}  

\title{Common femtoscopic hadron-emission source in \pp collisions at the LHC}
\ShortTitle{Exploring the source of mesons with Femtoscopy}   

\Collaboration{ALICE Collaboration\thanks{See Appendix~\ref{app:collab} for the list of collaboration members}}
\ShortAuthor{ALICE Collaboration} 

\begin{abstract}
The femtoscopic study of pairs of identical pions is particularly suited to investigate the effective source function of particle emission, due to the resulting Bose--Einstein correlation signal. In small collision systems at the LHC, \pp in particular, the majority of the pions are produced in resonance decays, which significantly affect the profile and size of the source. In this work, we explicitly model this effect in order to extract the primordial source in pp collisions at \thirteen from charged \spipi correlations measured by ALICE. We demonstrate that the assumption of a Gaussian primordial source is compatible with the data and that the effective source, resulting from modifications due to resonances, is approximately exponential, as found in previous measurements at the LHC. The universality of hadron emission in \pp collisions is further investigated by applying the same methodology to characterize the primordial source of \Kp pairs. The size of the primordial source is evaluated as a function of the transverse mass (\mt) of the pairs, leading to the observation of a common scaling for both \spipi and \Kp, suggesting a collective effect. Further, the present results are compatible with the \mt scaling of the \pP and p$-\Lambda$ primordial source measured by ALICE in high multiplicity pp collisions, providing additional evidence for the presence of a common emission source for all hadrons in small collision systems at the LHC. This will allow the determination of the source function for any hadron--hadron pairs with high precision, granting access to the properties of the possible final-state interaction among pairs of less abundantly produced hadrons, such as strange or charmed particles.

\end{abstract}
\end{titlepage}

\setcounter{page}{2} 

\input{Content/A_Introduction}
\input{Content/B_Data_Analysis}
\input{Content/C_Femtoscopy}
\input{Content/D_Modeling_of_the_source}
\input{Content/E_Results_and_discussion}
\input{Content/F_Summary}


\newenvironment{acknowledgement}{\relax}{\relax}
\begin{acknowledgement}
\section*{Acknowledgements}
\input{fa_2023-10-27_Opt_C.tex}
\end{acknowledgement}

\bibliographystyle{utphys}   
\bibliography{bibliography}

\newpage
\appendix
\input{Content/Z_Appendix}
\clearpage

\section{The ALICE Collaboration}
\label{app:collab}
\input{2023-10-27-Alice_Authorlist_2023-10-27_Opt_C.tex}
\end{document}

%% file: commands.tex
%

\newcommand{\pp}           {pp\xspace}
\newcommand{\ppbar}        {\mbox{$\mathrm {p\overline{p}}$}\xspace}
\newcommand{\XeXe}         {\mbox{Xe--Xe}\xspace}
\newcommand{\PbPb}         {\mbox{Pb--Pb}\xspace}
\newcommand{\pPb}          {\mbox{p--Pb}\xspace}
\newcommand{\AuAu}         {\mbox{Au--Au}\xspace}
\newcommand{\dAu}          {\mbox{d--Au}\xspace}

\renewcommand{\rm}         {\mathrm}

\newcommand{\roots}        {\ensuremath{\sqrt{s}}\xspace}
\newcommand{\snn}          {\ensuremath{\sqrt{s_{\mathrm{NN}}}}\xspace}
\newcommand{\pt}           {\ensuremath{p_{\rm{T}}}\xspace}
\newcommand{\meanpt}       {$\langle p_{\mathrm{T}}\rangle$\xspace}
\newcommand{\ycms}         {\ensuremath{y_{\rm CMS}}\xspace}
\newcommand{\ylab}         {\ensuremath{y_{\rm lab}}\xspace}
\newcommand{\etarange}[1]  {\mbox{$\left | \eta \right |~<~#1$}}
\newcommand{\yrange}[1]    {\mbox{$\left | y \right |~<~#1$}}
\newcommand{\dndy}         {\ensuremath{\mathrm{d}N_\mathrm{ch}/\mathrm{d}y}\xspace}
\newcommand{\dndpt}        {\ensuremath{\mathrm{d}N_\mathrm{ch}/\mathrm{d}p_{\rm T}}\xspace}
\newcommand{\dndeta}       {\ensuremath{\mathrm{d}N_\mathrm{ch}/\mathrm{d}\eta}\xspace}
\newcommand{\avdndeta}     {\ensuremath{\langle\dndeta\rangle}\xspace}
\newcommand{\dNdy}         {\ensuremath{\mathrm{d}N_\mathrm{ch}/\mathrm{d}y}\xspace}
\newcommand{\Npart}        {\ensuremath{N_\mathrm{part}}\xspace}
\newcommand{\Ncoll}        {\ensuremath{N_\mathrm{coll}}\xspace}
\newcommand{\dEdx}         {\ensuremath{\textrm{d}E/\textrm{d}x}\xspace}
\newcommand{\RpPb}         {\ensuremath{R_{\rm pPb}}\xspace}
\newcommand{\kstar}        {\ensuremath{k^*}\xspace}
\newcommand{\mt}           {\ensuremath{m_{\mathrm{T}}}\xspace}
\newcommand{\St}           {\ensuremath{S_{\mathrm{T}}}\xspace}
\newcommand{\kt}           {\ensuremath{k_{\mathrm{T}}}\xspace}
\newcommand{\ctau}         {\ensuremath{c\tau_\rm{res}}\xspace}
\newcommand{\ctaueq}[1]    {\mbox{$c\tau_\rm{res}=#1$}}
\newcommand{\ctaus}[1]     {\mbox{$c\tau_\rm{res}<#1$}}
\newcommand{\ctaul}[1]     {\mbox{$c\tau_\rm{res}>#1$}}
\newcommand{\ctaubtw}[2]   {\mbox{$#1<c\tau_\rm{res}<#2$}}
\newcommand{\ctaueff}   {\mbox{$\left\langle c\tau_\rm{res}^{\rm{eff}} \right\rangle$}\xspace}
\newcommand{\masseff}   {\mbox{$\left\langle m_\rm{res}^{\rm{eff}} \right\rangle$}\xspace}
\newcommand{\Xeff}   {\mbox{$\left\langle \mathcal{X}_\rm{res}^{\rm{eff}} \right\rangle$}\xspace}
\newcommand{\ctaueffeq}[1]   {\mbox{$\left\langle c\tau_\rm{res}^{\rm{eff}} \right\rangle~=~#1$}\xspace}
\newcommand{\masseffeq}[1]   {\mbox{$\left\langle m_\rm{res}^{\rm{eff}} \right\rangle~=~#1$}\xspace}

\newcommand{\nineH}        {$\sqrt{s}~=~0.9$~Te\kern-.1emV\xspace}
\newcommand{\seven}        {$\sqrt{s}~=~7$~Te\kern-.1emV\xspace}
\newcommand{\thirteen}     {$\sqrt{s}~=~13$~Te\kern-.1emV\xspace}
\newcommand{\twoH}         {$\sqrt{s}~=~0.2$~Te\kern-.1emV\xspace}
\newcommand{\twosevensix}  {$\sqrt{s}~=~2.76$~Te\kern-.1emV\xspace}
\newcommand{\five}         {$\sqrt{s}~=~5.02$~Te\kern-.1emV\xspace}
\newcommand{\twosevensixnn}{$\sqrt{s_{\mathrm{NN}}}~=~2.76$~Te\kern-.1emV\xspace}
\newcommand{\fivenn}       {$\sqrt{s_{\mathrm{NN}}}~=~5.02$~Te\kern-.1emV\xspace}
\newcommand{\LT}           {L{\'e}vy-Tsallis\xspace}
\newcommand{\GeVc}         {Ge\kern-.1emV/$c$\xspace}
\newcommand{\MeVc}         {Me\kern-.1emV/$c$\xspace}
\newcommand{\tev}          {Te\kern-.1emV\xspace}
\newcommand{\gev}          {Ge\kern-.1emV\xspace}
\newcommand{\mev}          {Me\kern-.1emV\xspace}
\newcommand{\GeVmass}      {Ge\kern-.1emV/$c^2$\xspace}
\newcommand{\MeVmass}      {Me\kern-.1emV/$c^2$\xspace}
\newcommand{\lumi}         {\ensuremath{\mathcal{L}}\xspace}

\newcommand{\ITS}          {\rm{ITS}\xspace}
\newcommand{\TOF}          {\rm{TOF}\xspace}
\newcommand{\ZDC}          {\rm{ZDC}\xspace}
\newcommand{\ZDCs}         {\rm{ZDCs}\xspace}
\newcommand{\ZNA}          {\rm{ZNA}\xspace}
\newcommand{\ZNC}          {\rm{ZNC}\xspace}
\newcommand{\SPD}          {\rm{SPD}\xspace}
\newcommand{\SDD}          {\rm{SDD}\xspace}
\newcommand{\SSD}          {\rm{SSD}\xspace}
\newcommand{\TPC}          {\rm{TPC}\xspace}
\newcommand{\TRD}          {\rm{TRD}\xspace}
\newcommand{\TZERO}        {\rm{T0}\xspace}
\newcommand{\VZERO}        {\rm{V0}\xspace}
\newcommand{\VZEROA}       {\rm{V0A}\xspace}
\newcommand{\VZEROC}       {\rm{V0C}\xspace}
\newcommand{\Vdecay} 	   {\ensuremath{V^{0}}\xspace}

\newcommand{\phot}         {\ensuremath{\gamma}\xspace} 
\newcommand{\ee}           {\ensuremath{e^{+}e^{-}}\xspace} 
\newcommand{\pip}          {\ensuremath{\uppi^{+}}\xspace}
\newcommand{\pim}          {\ensuremath{\uppi^{-}}\xspace}
\newcommand{\kap}          {\ensuremath{\rm{K}^{+}}\xspace}
\newcommand{\kam}          {\ensuremath{\rm{K}^{-}}\xspace}
\newcommand{\pbar}         {\ensuremath{\rm\overline{p}}\xspace}
\newcommand{\kzero}        {\ensuremath{{\rm K}^{0}_{\rm{S}}}\xspace}
\newcommand{\lmb}          {\ensuremath{\Lambda}\xspace}
\newcommand{\sigm}         {\ensuremath{\Sigma}\xspace}
\newcommand{\almb}         {\ensuremath{\overline{\Lambda}}\xspace}
\newcommand{\Om}           {\ensuremath{\Omega^-}\xspace}
\newcommand{\Mo}           {\ensuremath{\overline{\Omega}^+}\xspace}
\newcommand{\X}            {\ensuremath{\Xi^-}\xspace}
\newcommand{\Ix}           {\ensuremath{\overline{\Xi}^+}\xspace}
\newcommand{\Xis}          {\ensuremath{\Xi^{\pm}}\xspace}
\newcommand{\Oms}          {\ensuremath{\Omega^{\pm}}\xspace}
\newcommand{\degr}         {\ensuremath{^{\rm o}}\xspace}

\newcommand{\siZ}          {\ensuremath{\Sigma^{0}}\xspace}
\newcommand{\siP}          {\ensuremath{\Sigma^{+}}\xspace}
\newcommand{\siM}          {\ensuremath{\Sigma^{-}}\xspace}
\newcommand{\asiZ}         {\ensuremath{\overline{\Sigma^{0}}}\xspace}
\newcommand{\psiZCF}       {\rm{p}\mbox{--}\siZ}
\newcommand{\SNCF}         {\rm{N}\mbox{--}\ensuremath{\Sigma}\xspace}
\newcommand{\LNCF}         {\rm{N}\mbox{--}\ensuremath{\Lambda}\xspace}
\newcommand{\apasiZCF}     {\ensuremath{\overline{\mathrm{p}}}\mbox{--}\asiZ}
\newcommand{\psiZCombCF}   {\rm{p}\mbox{--}\siZ \ensuremath{\oplus} \apasiZCF}
\newcommand{\SB}           {\ensuremath{(\Lambda\gamma)}\xspace}
\newcommand{\pSBCF}        {\rm{p}\mbox{--}\SB}

\newcommand{\pLCF}         {\rm{p}\mbox{--}\lmb}
\newcommand{\nSplusCF}     {\rm{n}\mbox{--}\ensuremath{\Sigma^{+}}}

\newcommand{\Ledn}         {Lednick\'y\mbox{--}Lyuboshits approach\xspace}
\newcommand{\chiEFT}       {\ensuremath{\chi}\rm{EFT}\xspace}
\newcommand{\fss}       {\rm{fss2}\xspace}
\newcommand{\ESC}          {\rm{ESC16}\xspace}
\newcommand{\NSC}          {\rm{NSC97f}\xspace}

\newcommand{\Bbar}{\ensuremath{\mbox{B--$\mathrm{\bar{B}}}}\,}


\newcommand{\aprot}{\ensuremath{\mathrm{\bar{p}}}\,}
\newcommand{\prot}{\ensuremath{\mathrm{p}}\,}
\newcommand{\pP}{\ensuremath{\mathrm {p\mbox{--}p}}\,}
\newcommand{\pXi}{\ensuremath{\mathrm {p\mbox{--}\Xi}}\,}
\newcommand{\pOmega}{\ensuremath{\mathrm {p\mbox{--}\Omega}}\,}
\newcommand{\LXi}{\ensuremath{\mathrm {\Lambda\mbox{--}\Xi}}\,}
\newcommand{\phiP}{\ensuremath{\mathrm {\phi\mbox{--}p}}\,}
\newcommand{\akP}{\ensuremath{\mathrm {K^-\mbox{--}p}}\,}
\newcommand{\Kp}{\ensuremath{\mathrm {K\mbox{--}p}}\,}
\newcommand{\aKp}{\ensuremath{\mathrm {K^+\mbox{--}p}}\,}
\newcommand{\Kap}{\ensuremath{\mathrm {K^-\mbox{--}\overline{p}}}\,}
\newcommand{\pipi}{\ensuremath{\mathrm {\uppi^+\mbox{--}\uppi^+}}\,}
\newcommand{\apiapi}{\ensuremath{\mathrm {\uppi^-\mbox{--}\uppi^-}}\,}
\newcommand{\spipi}{\ensuremath{\mbox{$\uppi$--$\uppi$}}~}
\newcommand{\pap}{\ensuremath{\mathrm {p\mbox{--}\bar{p}}}\,}
\newcommand{\paL}{\ensuremath{\mathrm {p\mbox{--}\overline{\Lambda}}}\,}
\newcommand{\apL}{\ensuremath{\mathrm {\overline{p}\mbox{--}\Lambda}}\,}
\newcommand{\LaL}{\ensuremath{\mathrm {\Lambda\mbox{--} \overline{\Lambda}}}\,}
\newcommand{\pL}{\ensuremath{\mathrm {p\mbox{--}\Lambda}}\,}
\newcommand{\ApaL}{\ensuremath{\mathrm {\bar{p}\mbox{--}\overline{\Lambda}}}\,}
\newcommand{\BBbar}{\ensuremath{\mathrm {B\mbox{--} \bar{B}}}\,}
\newcommand{\BB}{\ensuremath{\mathrm {B\mbox{--} B}}\,}
\newcommand{\CF}        {\ensuremath{C(k^*)}\xspace}

\newcommand{\rs}           {\ensuremath{r^*}\xspace}
\newcommand{\rsv}           {\ensuremath{\vec{r}^*}\xspace}
\newcommand{\rc}           {\ensuremath{r_\mathrm{core}}\xspace}
\newcommand{\rcv}           {\ensuremath{\vec{r}_\mathrm{core}}\xspace}
\newcommand{\rcs}           {\ensuremath{r^*_\mathrm{core}}\xspace}
\newcommand{\rcsv}           {\ensuremath{\vec{r}^*_\mathrm{core}}\xspace}
\newcommand{\ks}           {\ensuremath{k^*}\xspace}
\newcommand{\ksv}           {\ensuremath{\vec{k}^*}\xspace}
\newcommand{\Sr}           {\ensuremath{S(r)}\xspace}
\newcommand{\Ck}           {\ensuremath{C(k)}\xspace}
\newcommand{\Srs}           {\ensuremath{S(\rs)}\xspace}
\newcommand{\Cks}           {\ensuremath{C(\ks)}\xspace}

\newcommand{\Rplus}{\protect\hspace{-.1em}\protect\raisebox{.35ex}{\smaller{\smaller\textbf{+}}}}
\newcommand{\Cpp}{\mbox{C\Rplus\Rplus}\xspace}

%% file: Content/A_Introduction.tex
\section{Introduction}

Femtoscopy is a technique of connecting the pair-wise momentum correlations to the properties of particle emission and the subsequent final-state interaction (FSI)~\cite{KooninPratt,Fabbietti:2020bfg}. This formalism is often applied to identical charged pions in order to study the effective source function of particle emission. This is achieved by taking advantage of the Bose--Einstein correlation signal present in the same-sign \spipi system. These techniques are quite advanced in heavy-ion collisions, for which the source function is well described by a Gaussian parameterization and the corresponding source size shows a distinct decrease as a function of the pair transverse mass (\mt). The latter is commonly referred to as \mt scaling and typically attributed to the presence of collective phenomena, i.e. radial flow~\cite{KooninPratt}, and is well described by models with a hydrodynamic phase~\cite{THERMINATOR2, THERMINATOR2piKp, iHKM}. Recently, femtoscopic studies in small collision systems, \pp in particular, have gained a lot of attention due to the possibility of producing pairs with low relative momentum at distances comparable to the range of the strong force ($\approx~1$--$2~$fm). The signal related to the FSI, embedded in the measured correlation function, allows for testing of theoretical predictions as well as extracting information about the low-energy scattering parameters, such as the scattering length and effective range. Nevertheless, the successful application of this technique requires the source function to be well constrained. The ALICE Collaboration has demonstrated that a common emission of primordial baryons, i.e. particles produced during the initial collision, is present in pp collisions by analyzing \pP and \pL correlations~\cite{CommonSource}. Similar to heavy-ion collisions, the results in \pp collisions show a clear \mt scaling, the nature of which is not yet understood. While hydrodynamic models are well motivated and established for heavy-ion collisions~\cite{ALICE:review}, this is not the case for pp collisions where there is controversy on the matter. The modeling of the source is realized with the so-called Resonance Source Model (RSM)~\cite{CommonSource}, which assumes a universal Gaussian core for all primordial particles and accounts for hadron production through resonances, finally yielding an effective source. The abundances of resonances are obtained from the statistical hadronization model~\cite{Andronic:2017pug}, assuming a canonical ensemble, while the specific decay kinematics can be extracted from a transport model, e.g.~EPOS~\cite{EPOS}. This opened the possibility to test for the first time both lattice QCD predictions for the more rare pairs \pXi and \pOmega~\cite{pOmega}, to probe the coupled-channel dynamics for \akP~\cite{ALICEKaonProton}, to constrain the chiral effective field theory using the \pL~\cite{pL} and \LXi~\cite{LXi} channels, and to access the interaction of a vector meson with baryons through \phiP~\cite{Phip}. Although the \mt scaling is not explicitly modeled within the RSM, an independent analysis employing a new framework, called CECA~\cite{CECA}, relates the \mt scaling to the radial expansion of the collision system.

The goal of this work is to explore if the ansatz of a common source can be extended to the meson$-$meson and meson$-$baryon sectors. This is achieved by employing the RSM to extract the effective Gaussian size of the same-sign primordial \spipi~(\pipi and \apiapi) and \Kp~(\aKp and~\Kap) sources as a function of \mt, and investigate if they are compatible with the \pP source extracted from the same data set collected from high-multiplicity \pp collisions at \thirteen. In this way, the assumption of a universal hadron emission in small collision systems will be solidified. Furthermore, as the found \mt scaling is currently not understood, the additional information on the source size at very low-\mt ($< 1$~GeV/$c^2$) will provide valuable information for transport~\cite{EPOS, PYTHIA} and effective source~\cite{CECA} models. Finally, we perform the \spipi correlation analysis in minimum-bias (MB) events using three different multiplicity classes in order to provide even more differential data.

The article is structured as follows. In Sec.~\ref{sec:dataanalysis} details regarding the data analysis are given, in particular about the particle identification using the ALICE detector. This is followed by a brief review of the femtoscopic method in Sec.~\ref{sec:Femto}, which includes the model used for fitting the experimental correlation functions. The source function for the studied pairs is discussed in Sec.~\ref{sec:Resonances}, which entails an investigation on the modification of the source function due to short-lived resonances. Section~\ref{sec:results} focuses on the obtained results, which are then discussed within the context of recent measurements. Finally, a summary is given in the closing Sec.~\ref{sec:conclusion}. 

%% file: Content/B_Data_Analysis.tex
\section{Data analysis}\label{sec:dataanalysis}
The data analyzed in this work were collected by the ALICE Collaboration~\cite{ALICEDetector} at the LHC in \pp collisions at \thirteen.
Additional details on the ALICE detector setup can be found in~\cite{ALICEperf}.
In this analysis, the following main subdetectors are used: the $\VZERO$ detectors~\cite{ALICEVZERO}, the Inner Tracking System (ITS)~\cite{ALICEITS}, 
the Time Projection Chamber (TPC)~\cite{ALICETPC}, and the Time-Of-Flight (TOF) detector
\cite{ALICETOF}.
Events passing a MB trigger are selected by requiring coincident signals in both $\VZERO$~\cite{ALICEVZERO} detectors to be synchronous with the beam crossing time defined by the LHC clock. High-multiplicity (HM) events are selected online by applying a threshold on the amplitude of the signal in the V0 detectors in addition to the MB trigger conditions.
The $\VZERO$ detectors consist of two arrays of plastic scintillators, one located at forward ($2.8<\eta<5.1$) and the other placed at backward ($-3.7<\eta<-1.7$) pseudorapidity~\cite{ALICEVZERO}. The measured amplitudes within these scintillators are used to define multiplicity classes for events. Additional measurements at midrapidity, utilizing the innermost layers of the ITS, then provide estimates for the average charged particle multiplicity at mid-pseudorapidity ($\eta$=0) in a unit interval of pseudorapidity $\langle\mathrm{d}N/\mathrm{d}\eta\rangle_{|\eta|<0.5}$ in a given multiplicity class. This procedure allows one to associate $\langle\mathrm{d}N/\mathrm{d}\eta\rangle_{|\eta|<0.5}$ with a V0-based multiplicity class.
For the HM case, the selected events correspond to the 0.17\% highest multiplicity inelastic \pp collisions with at least one charged particle in the range $|\eta| < 1$ (referred to as INEL $>$ 0)~\cite{ALICEDetector,ALICEperf}. 
The resulting HM data sample contains events with an average multiplicity of approximately 30 charged particles in the pseudorapidity interval \etarange{0.5}~\cite{pOmega}.
Events with multiple primary vertices, identified from track segments in the two innermost layers of the ITS, are tagged as pile-up and removed from the analyzed sample.

The charged-particle tracking and the primary vertex (PV) reconstruction are obtained using the combined track information of the ITS and TPC~\cite{ALICEDetector}, located inside a uniform magnetic field  of 0.5~T directed along the beam direction. A selection of $\pm10$ cm on the deviation between the \textit{z} coordinate of the reconstructed PV and the nominal interaction point is applied to ensure a uniform detector coverage in the pseudorapidity region of $|\eta|<0.8$. The quality of the tracks used for the construction of the \spipi (\Kp) correlation is ensured by requiring that each track lies within the pseudorapidity range of \etarange{0.8} while also requiring that a minimum of $80$~($70$) clusters in the TPC~\cite{ALICETPC} are assigned to each track. Further quality assurance is provided by requiring that at least 70 out of the total of 159 rows of the TPC readout pads are crossed in conjunction with finding clusters in at least 80\% of these crossed rows. For this analysis, different transverse momentum ranges for the tracks were considered depending on the particle species and the particle-identification (PID) capabilities of ALICE. For pions, protons and kaons these ranges are $0.14 < \pt < 4 $~\GeVc, $0.8 < \pt < 3$~\GeVc, and $0.4 < \pt < 1.4$~\GeVc, respectively.
The PID for the charged pions, kaons, and protons is performed using the TPC and TOF detectors. The deviation of the measured TPC response to the Bethe-Bloch parameterization of $\langle \mathrm{d}E/\mathrm{d}x\rangle$ for the mass hypothesis of the considered particle is expressed as multiples of the standard deviation $\mathrm{n}\sigma$, determined by the resolution of the TPC.
For pions, only those tracks satisfying $|n_{\sigma,\TPC}| < 3$ for a transverse momentum $\pt < 0.5$~\GeVc are further analyzed. In the range $\pt > 0.5$~\GeVc, the pion specific
energy loss band overlaps with those of kaons and protons. Hence, up to $\pt < 4$~\GeVc, also the TOF information is used. The difference between the measured TOF response and the velocity calculated for a pion as a function of the momentum is expressed as $n_{\sigma,\TOF}$, and combined with the TPC information into $n_{\sigma,\mathrm{combined}}=(n^2_{\sigma,\TPC}+n^2_{\sigma,\TOF})^{(1/2)}$. Only tracks satisfying $|n_{\sigma,\mathrm{combined}}| < 3$ are further analyzed.
The selection for charged kaons (protons) is described in~\cite{ALICEKaonProton} and follows the same strategy for PID used for the charged pions. For $0.15~(0.4) < \pt < 0.4~(0.8)$~\GeVc only $\TPC$ information is used, and the $|n_{\sigma,\TPC}| < 3$ criterion is applied. Again, the combined $\TPC$ and $\TOF$ information is used for $0.4~(0.8) < \pt < 1.4~(3)$~\GeVc and required to be less than $3$ times of $n_{\sigma,\mathrm{combined}}$. In order to remove a large fraction of the e$^{\pm}$ contamination of the $\langle \mathrm{d}E/\mathrm{d}x\rangle$ bands of charged kaons (protons), the identified particle candidates were excluded in the region of $0.3~(0.6) < \pt < 0.4~(0.8)$~\GeVc. As it was shown in Refs.~\cite{ALICEKaonProton,ALICEKpInppNew}, this selection does not introduce any bias to the correlation function.

For femtoscopic studies, the primary particles, which experience the final-state interaction, are of special interest, and the amount of secondary (feed-down) particles stemming from weak decays and interactions in the detector material have to be suppressed. This is achieved by requiring that the distance-of-closest approach (DCA) to the primary vertex of the tracks used for the \spipi (\Kp) correlation is less than $0.3$~($1.0$)~cm in the transverse and longitudinal planes with respect to the beam direction.

The two-particle correlation function is measured as a function of the relative momentum \newline $(\ks~=~\frac{1}{2}|\Vec{p}_1^*~-~\Vec{p}_2^*|)$, where all quantities, denoted with an asterisk~(*), are evaluated in the rest frame of the particle pair (PRF). The momentum in the PRF of particle $i$ is denoted by $\Vec{p}_i^*$.
The measured correlation function contains in principle all correlations imprinted on the relative momentum (\ks) distribution regardless of the origin, consequently the background has to be taken into account. This correlated residual background stems from hard partonic interactions in the initial stage of the collision, and leads to highly correlated particles in jet-like structures, leading typically to less isotropic events. For the analysis of the correlation functions of \spipi pairs from MB events and of \Kp pairs from HM events, a selection based on the event shape observable called transverse sphericity~\cite{SphericityDefinition, Sphericity} \St$>0.7$ is applied in order to suppress the mini-jet background, largely present in \pp collisions~\cite{Sphericity,ALICEKaonProton,ALICELKpp,ALICEpipiInpp,BantiB,Phip}. No selection on the sphericity is applied in the case of \spipi correlations for the events in the HM sample because these events are typically more spherical, and because the selection of $\St> 0.7$ affects the relative momentum correlation function of \spipi pairs, mostly for $\ks> 500$~\MeVc, which is well outside the signal region of $\ks< 250$~\MeVc. 
After applying the event selection criteria, the analyzed sample consists of about $10^9$ MB and $10^9$ HM events. 

For tracks close in phase space, reconstruction errors can occur in the form of track merging or splitting~\cite{ALICECPR}. Therefore, an additional selection for the close-pair rejection is included by imposing a circular cut on the relative angular separation measured by the relative pseudorapidity $\Delta\eta$ and azimuthal angle $\Delta\varphi^{*}$, $\sqrt{\Delta\varphi^{*2} + \Delta\eta^2} < (1\times10^{-2})$ of the tracks, hence, demanding a minimum separation. Here $\Delta\varphi^{*}$ is corrected for the change in azimuthal angle induced by the magnetic field. The use of this selection completely removes the signatures of track merging or splitting.

The purity of the PID selections and the contamination due to weak decays (feed-down) of the sample were studied with the help of Monte Carlo (MC) simulations using PYTHIA 8.2~\cite{PYTHIA} (Monash 2013 Tune) generated events, which were transported by GEANT 3~\cite{GEANT} through the ALICE detector and subsequently processed by the reconstruction algorithm~\cite{ALICEDetector}. The purity has been studied as a function of \pt, nevertheless a negligible dependence, even across the threshold at which the TOF information is used, was observed. The \pt averaged purity obtained with the PID selections described above is about $99\%$ for charged pions, kaons, and protons. The primary and secondary fractions are obtained by extracting from the MC the template distributions of the DCA in the transverse plane and fitting these to the measured distributions.
The information about the purity and composition of the sample is crucial for femtoscopic studies and is used to determine the $\lambda$ parameters, which are introduced in Sec.~\ref{sec:Femto}.

The correlation function is studied differentially in intervals of the pair transverse mass \mt, which is connected to the average transverse momentum of the particle pair ($\kt = \frac{1}{2}|\Vec{\pt}_1 + \Vec{\pt}_2|$ ) via the relation $\mt = (\kt^2 + m_{\mathrm{avrg}}^2)^{(1/2)}$, where $m_{\mathrm{avrg}}$ is the average mass of the studied particle pair.
The \spipi correlations are studied in five (0.15--0.30, 0.30--0.50, 0.50--0.70, 0.70--0.90, 0.90--1.50 \GeVc) \kt ranges for the MB and HM datasets. The \Kp correlations are studied in four (1.2--1.4, 1.4--1.5, 1.5--1.8, 1.8--2.0 \GeVmass) \mt ranges in HM collisions. 

The systematic uncertainties of the correlation function associated with the choice of the selection criteria are estimated by performing the selection 44 times while varying randomly the track selection and PID criteria (i.e.~\pt boundaries at which TOF is used and the $n\sigma$ threshold) by $20$\% and excluding the \St selection from the procedure. The choice of including no variation on \St was made to ensure that the same events are always used as input in the analysis. In order to ensure that the observed deviation with respect to the default settings does not originate from statistical fluctuations, only variations for which the yield of pairs for $\ks<200$~\GeVc does not exceed $20$\% compared to the yield obtained from the default settings are used. The magnitude of the uncertainty is estimated using the root-mean square deviation for a uniform distribution in each \ks interval, using the underlying uniform distribution obtained from the 44 variations.

%% file: Content/C_Femtoscopy.tex
\section{Two-particle correlations}
\label{sec:Femto}
The observable of interest in femtoscopic studies is the two-particle correlation function \CF, measured as a function of \ks. Experimentally, \CF is calculated from the normalized ratio

\begin{align}\label{eq:CFexp}
\CF & = \mathcal{N} N_{\mathrm{same}}(\ks)/N_{\mathrm{mixed}}(\ks),
\end{align}

where $N_{\mathrm{same}}(\ks)$ is the distribution of \ks obtained from particles produced in the same event, and $N_{\mathrm{mixed}}(\ks)$ is the \ks distribution of uncorrelated pairs. The latter is generated using the mixed event technique~\cite{KooninPratt}, for which a pair is built by selecting particles from different events. To avoid any bias pertaining to reconstruction efficiencies or acceptance, only identified particles are used from events for which the difference in the $z$ coordinate of the reconstructed primary vertex is less than $2$~cm and the difference in \dNdy at midrapidity ($|y| <0.5$) is less than 4. The normalization $\mathcal{N}$ is calculated in $\ks~\in~[350,400]$~\MeVc, in which no femtoscopic signal is present, and the theoretical value of \CF approaches unity. In order to extract the source parameters a femtoscopic fit is performed using a model for the expected theoretical correlation function.

The theoretical definition of \CF is given by the Koonin--Pratt equation~\cite{KooninPratt}

\begin{align}\label{eq:CFKooninPratt}
\CF~=~\int \mathrm{d^3}\rs S(\rs) |\psi(\rs,\ks)|^{2}, 
\end{align}

and depends on the source function $S(\rs)$, which encodes the relative distance \rs in the PRF, and the two-particle relative wave function $\psi(\rs,\ks)$. In this work, the latter is obtained by employing the "Correlation Analysis Tool using the Schrödinger Equation" (CATS) framework~\cite{CATS}, which numerically solves the Schrödinger equation for a configurable interaction potential. In the case of the \spipi pairs, the wave function is determined by quantum statistics and Coulomb repulsion. For the \Kp correlations, the strong interaction is modeled assuming state-of-the-art chiral potentials~\cite{ChiralKplusp, ALICEKpInppNew}, while the Coulomb interaction is directly taken into account using CATS. 
The shape of $S(\rs)$ follows a convolution of a Gaussian distribution (core) and an exponential function (tail)~\cite{CommonSource}. The Gaussian core, from which primordial particles are emitted, is fully determined by its width \rc and the exponential contribution takes into account the resonance decays to the pair of interest. Further details on the $S(\rs)$ are discussed in Sec.~\ref{sec:Resonances}.
The Koonin--Pratt relation is a subject to several approximations, as discussed in~\cite{KooninPratt}. These approximations have been originally designed for large emission sources, as found in heavy-ion collisions, nevertheless they have been successfully applied in identical pion femtoscopy in \pP collision systems~\cite{ALICEpipiInpp, CMSpipiInpp, ATLASpipi}. 
 
\sloppy The measured \CF encompasses not only correlations arising from primordial pairs but also incorporates contributions from weak decays, long-lived resonances (with a proper decay length greater than 5~fm) and misidentified particles. These additional components consist of what are commonly referred to as "feed-down" particles and impurities, and they can be taken into account by the $\lambda$ parameter formalism, as was shown in~\cite{LambdaParams}. The general form of the modeled correlation function $C_{\mathrm{model}}(\ks)$ is given by

\begin{equation}
    C_{\mathrm{model}}(\ks) = 1 + \sum_i \lambda_i (C_{\mathrm{}}(\ks) -1).
\label{eq:FemtoGeneral}
\end{equation}

In this analysis, a decomposition of the correlation function into three components is considered, namely the genuine (gen), the feed-down (feed), and the misidentified candidates (misid), leading to 

\begin{equation}
    C_{\mathrm{model}}(\ks) = 1 + \lambda_{\mathrm{gen}}(C_{\mathrm{gen}}(\ks) -1) +\lambda_{\mathrm{feed}} (C_{\mathrm{feed}}(\ks) -1) +\lambda_{\mathrm{misid}} (C_{\mathrm{misid}}(\ks) -1).
\label{eq:FemtoPion}
\end{equation}

By definition, the sum of the three contributions is normalized to unity~\cite{LambdaParams}. 
The genuine contribution contains the pairs of primary particles experiencing the final-state interaction and particles stemming from short-lived strongly decaying resonances. Details about the considered resonances are given in the following Sec.~\ref{sec:Resonances}. The pairs containing at least one particle originating from weak or electromagnetic decays are described by the feed-down component. Finally, the pairs with at least one misidentified particle are quantified by the misidentification part. The latter additionally includes particles which are produced in interactions between the particles produced in the event and the detector material, whose overall contribution is however negligible. The \CF associated with the feed-down and misidentified components are considered to be flat, leading to a simplified form of Eq.~\eqref{eq:FemtoPion}:

\begin{equation}
    C_{\mathrm{model}}(\ks) = \lambda_{\mathrm{gen}}C_{\mathrm{gen}}(\ks) +\lambda_{\mathrm{feed}} +\lambda_{\mathrm{misid}}.
\label{eq:FemtoPionsimple}
\end{equation}

The $\lambda_i$ parameters are evaluated from the \mbox{single-particle} properties as a product of the purity~$\mathcal{P}_i$, determined via MC studies and fractions of primary particles~$f_i$ obtained from fits to the experimental DCA distributions in the transverse plane. The formula used for the evaluation is $\lambda_{ij} = \mathcal{P}_if_i\mathcal{P}_jf_j$, where the indices $i$ and $j$ indicate the origin (in this case primary, feed-down and misidentification) of the single particles which form the pair. The $\lambda_{\mathrm{gen}}$ and $\lambda_{\mathrm{feed}}$ are determined as described above, $\lambda_{\mathrm{misid}}$ is then obtained by the constraint that the sum of the three contributions is normalized to unity, and driven by the purity.

To account for any residual non-femtoscopic  background, the function used in the femtoscopic fit to the measured \CF is modified by an additional baseline

\begin{align}
C_{\mathrm{fit}}(\ks) = B_{\mathrm{non-femto}}(\ks) \times C_{\mathrm{model}}(\ks),
\end{align}
where $B_{\mathrm{non-femto}}(\ks)$ is the baseline function of the residual non femtoscopic background. 
In the case of same-sign pions, two functions for the baseline are tested, a polynomial of first and second degree, while for the \Kp correlations a constant, a first, and a second order polynomial are considered. In \spipi correlations from the MB sample, a residual mini-jet background is also present after applying the sphericity selection. Hence, to remove this remaining non-femtoscopic component, the \CF obtained from MB data is divided by the correlation function obtained from MC before using it in the femtoscopic fit. The same strategy was already employed by the ATLAS and CMS Collaborations~\cite{CMSpipiInpp, ATLASpipi}. The typical size of this correction is around 10\%. The free parameters of the fit are the baseline parameters, and the width \rcs of the source function. The fit is performed in the range of $0<\ks<364 (270)$ \MeVc for \spipi (\Kp). A variation of $\pm 10\%$ in the upper limit of the fitting range is considered to account for any systematic effect related to the range chosen for the fit.

%% file: Content/D_Modeling_of_the_source.tex
\section{Modeling of short-lived resonances}
\label{sec:Resonances}
The detailed study of the particle source function $S(\rs)$ for baryon$-$baryon pairs was presented in~\cite{CommonSource} and enabled studies of the interaction between many particle pairs, including for example, \pOmega~\cite{pOmega}. In this work, the MC procedure developed in~\cite{CommonSource} serves as a baseline to determine, for the first time, the emitting source for primordial particles as a function of \mt for \mbox{meson$-$meson} and \mbox{meson$-$baryon} pairs.

The source function $S(\rs)$ according to the RSM~\cite{CommonSource} is a two component ansatz. A Gaussian source is used for primordially produced particles, i.e.~initial particles produced in the \pp collisions. It has a  width $\rc$, called ``core source''. The general Gaussian parameterization for the two-particle source is

\begin{equation}
    S(\rs) = \frac{1}{(4\pi r_0^2)^{3/2}}\exp{\left(-\frac{r^{*2}}{4r_0^2}\right)},
\label{eq:FemtoSourceGauss}
\end{equation}

this form is obtained by computing the convolution of two single particle Gaussian sources, each with a respective width of $r_0$. 
In addition to this, the contributions of
short-lived strongly-decaying resonances can be parameterized by introducing exponential tails, the so-called ``resonance halo''. The analytical determination of the full shape of the source is challenging, because of the kinematics of decaying resonances. Therefore, a MC procedure is employed to access numerically the source function, which can be parameterized by folding the Gaussian in Eq.~\eqref{eq:FemtoSourceGauss} with the sum of the exponential decay functions for every resonance. The decay kinematics is obtained by extracting the angular distribution of particles emitted from the strong decay of the short-lived resonances from the EPOS~\cite{EPOS} model version 3.117, which is configured to simulate MB \pp collisions at \thirteen with relativistic hydrodynamics and hadronic UrQMD afterburner enabled. In previous studies~\cite{ATLASpipi,CMSpipiInpp,CMSpipiInAA}, the effect of resonances was not explicitly taken into account, instead a Cauchy/Exponential type source parameterization~\cite{Csorgo2004} was used

\begin{equation}
    S(\ensuremath{r^*}\xspace) = \frac{1}{\pi^2}\frac{r_{\mathrm{exp}}}{{\left(r_{\mathrm{exp}}^2 +r^{*2}\right)}^{2}}.
\label{eq:FemtoSourceCauchy}
\end{equation}
where the size of the source is denoted by $r_{\mathrm{exp}}$.

Within the RSM, the abundances of strongly decaying resonances contributing to the yield of a specific particle are constrained by calculations based on a canonical implementation of the statistical hadronization model (SHM)~\cite{ThermalFIST}. For charged pions, kaons, and protons, the \textsc{Thermal-FIST}~\cite{ThermalFIST} package is employed to perform the calculations. The model parameters are evaluated following the procedure described in~\cite{ThermalFISTparam}, separately for MB and HM collisions. According to these calculations, around $28\%$ of the charged pions are primordial, while $72\%$ stem from strongly decaying resonances. For charged kaons (protons) the primordial fraction is $52~(36)\%$. The resonances which contribute at least 1\% to the yield of the charged pions and kaons are listed in Tables~\ref{tab:ThermalFIST_resoYields_chargedpions} and~\ref{tab:ThermalFIST_resoYields_chargedkaons}, respectively, while the detailed decomposition of the proton resonances can be found in Ref.~\cite{CommonSource}. The mass and lifetime of each strongly decaying resonance are embedded in two effective parameters \masseff and \ctaueff, which are the average mass and decay length computed using as weights the abundances and lifetimes of the considered resonances. As the difference in the relative contribution of resonances is negligible between MB and HM events, in both cases, the same values for the effective parameters are used. The parameters used for the calculations are reported in Table~\ref{tab:ThermalParam} for the MB collisions. 
\begin{table}
\begin{center}    
\caption[Thermal-FIST configuration]{Model specifications and values of parameters used for the yield calculations with \textsc{Thermal-FIST}~\cite{ThermalFIST}.} 
\begin{tabular}{l | c}
Ideal HRG model specifications\\
\hline \hline
Ensemble & Canonical \\
Statistics & Quantum statistics for all particles \\
Resonance shape & rel. Breit-Wigner distribution\\
\hline
Parameter & Value \\
\hline\hline
Temperature (MeV) &  171.0 \\
Strangeness suppression factor & $\gamma_S = 0.78$ \\
Source radius for $\mid y \mid<1$ (fm) & $R = 1.58$ \\
Canonical correlated radius (fm) & $R_\rm{c} = 2.28$ \\
B, Q, S & 0\\
\end{tabular}
\label{tab:ThermalParam}
\end{center}
\end{table}
The results of these calculations are \masseffeq1.124~\GeVmass and \ctaueffeq1.5~fm for the charged pions and \masseffeq1.05 (1.36)~\GeVmass and \ctaueffeq3.66 (1.65)~fm for the charged kaons (protons). Long-lived ($c\tau > 5$~fm) resonances are excluded from the calculation of these average values, as explained in the next paragraph. Along with the decay kinematics obtained by selecting a particle cocktail of matching \masseff and \ctaueff from events generated with EPOS, all ingredients needed for the RSM are determined. Due to the averaging of the resonance properties it is possible that especially for low \kt pion pairs the \masseff and \ctaueff are overestimated. However, this is accommodated by including a systematic variation of both parameters. A variation of $\pm 10\%$ for \masseff and \ctaueff is considered for the fits to account for any systematic effects related to the hadronic cocktail obtained from the SHM as well as the averaging. This procedure may be improved upon, in the future, by simulating each decay independently; first steps in this direction become possible with a new framework handling the particle emission called CECA~\cite{CECA}.

\begin{table}[!ht]
\begin{center}    
\caption{List of resonances which contribute at least 1\% to the yield of \pip, the contribution of the remaining resonances is summarized by $R_{f<1\%}^{\pi}$ The fractions are computed with \textsc{Thermal-FIST} for \pp minimum bias collisions at \thirteen and are also used for the high multiplicity sample since the relative abundance of resonances is the same as for minimum bias collisions.}

\begin{tabular}{|l |c|}
\hline
Resonances & Fraction (\%) \\
\hline 
$\rm{\rho(770)^0}$ & {9.0} \\
$\rm{\rho(770)^+}$ & {8.7} \\
$\rm{\omega}$({782}) & {7.7} \\
$\rm{K^{*}({892})^+}$  & {2.3} \\
$\rm{\overline{K}^*({892})^0}$  & {2.6} \\
$\rm{b_1({1235})^0}$  & {1.9} \\
$\rm{a_2({1320})^+}$  & {1.5} \\
$\rm{\eta}$ & {1.5} \\
$\rm{a_1({1260})^+}$  & {1.4} \\
$\rm{f_2({1270})}$  & {1.4} \\
$\rm{a_0({980}) ^+}$ & {1.4} \\
$\rm{h_1({1170})}$  & {1.2} \\
$R_{f<1\%}^{\pi}$  & {31.4} \\
\hline
\end{tabular}
\label{tab:ThermalFIST_resoYields_chargedpions}
\end{center}
\end{table}

\begin{table}[!ht]
\begin{center}    
\caption{List of resonances contributing at least 1\% to the yield of $\rm{K^{+}}$, the contribution of the remaining resonances is summarized by $R_{f<1\%}^{K}$. These fractions are computed with \textsc{Thermal-FIST} for \pp high multiplicity collisions at \thirteen.}

\begin{tabular}{|l |c|}
\hline
Resonances & Fraction ($\%$) \\
\hline 
$\rm{K^*(892)^0}$ & 21.0 \\
$\rm{K^*(892)^+}$ & 11.0 \\
$\rm{a_0(980)^+}$ & 1.0 \\
$\rm{K^{*}_{2}(1430)^0}$ & 1.0 \\
$\rm{K^{*}_{1}(1270)^0}$ & 1.0 \\
$\phi$ (1020)  & 6.0 \\
$R_{f<1\%}^{K}$  & {8.0} \\\hline
\end{tabular}
\label{tab:ThermalFIST_resoYields_chargedkaons}
\end{center}
\end{table}

\begin{figure}[ht!]
\centering
\includegraphics[width=.47\textwidth]{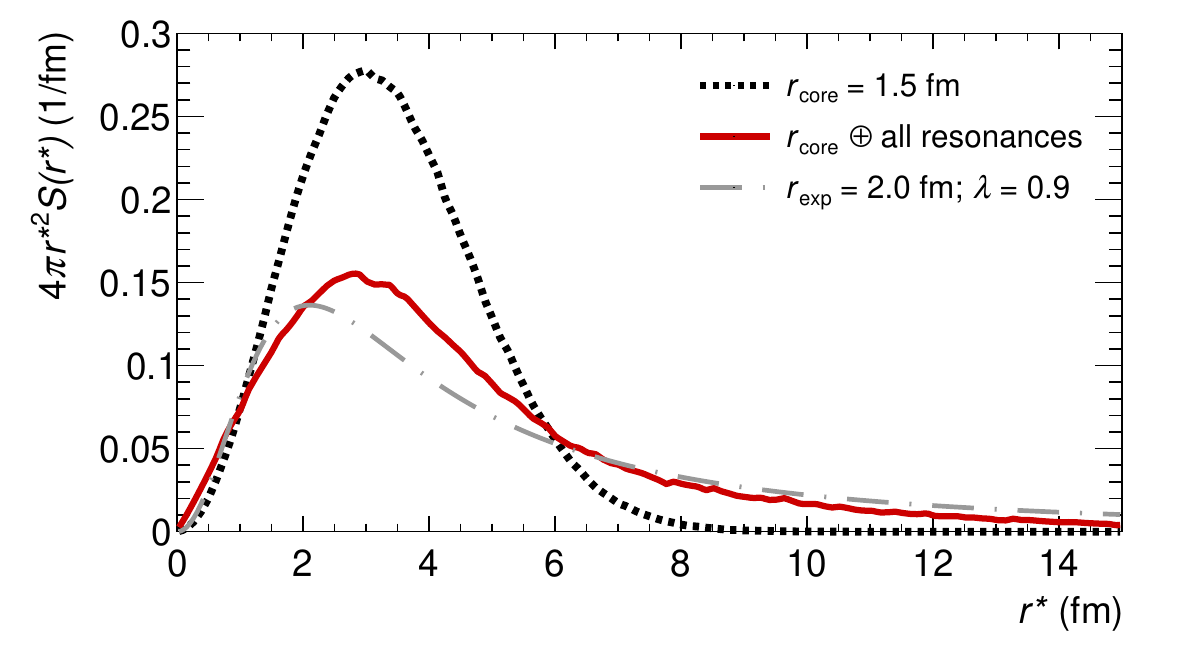}
\includegraphics[width=.47\textwidth]{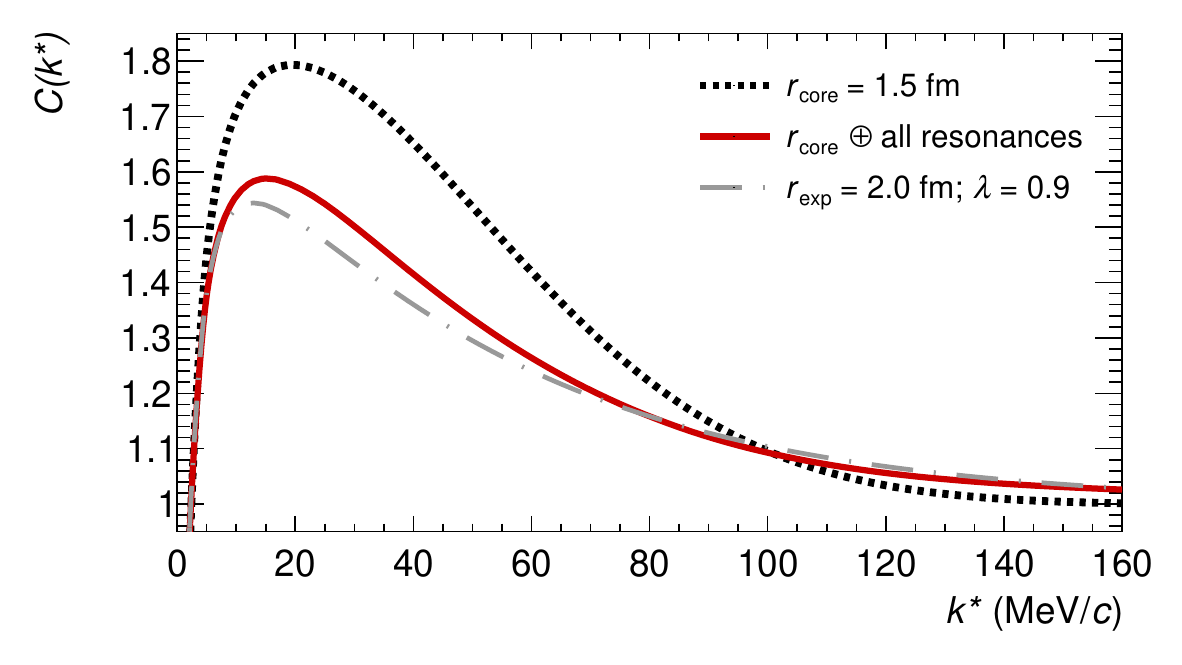}
\caption{Calculation of the source function (left) and corresponding \spipi correlation function (right), performed using CATS~\cite{CATS} employing the RSM. The RSM~(full red line) is compared to the usual assumption of an exponential~(dash-dotted grey line) source distribution with a radius $r_{\mathrm{exp}}$ of the source and a purely Gaussian~(dotted black line) source.}
\label{fig:RSM_Sr_and_Ck}
\end{figure}

The result of the RSM for the full \spipi source profile, together with the Gaussian core, and an exponential type source profile is shown in the left panel of Fig.~\ref{fig:RSM_Sr_and_Ck}, while the resulting correlation functions are depicted in the right panel. Comparing the purely Gaussian source, as well as the corresponding correlation function, with the scenario of a source containing exponential tails shows that resonance contributions diminish the overall correlation strength. The Gaussian core source together with the resonance contributions (solid red curve), can effectively be described by an exponential source, scaled by $\lambda = 0.9$ due to contributions of $\approx 10\%$ of long-lived ($c\tau > 5$~fm) strongly decaying resonances (dash-dotted grey curve). Note that $\lambda = 0.9$ can conveniently be absorbed by the usual $\lambda$ parameters introduced in Eq.~\eqref{eq:FemtoPion}, especially in the case of a simultaneous fit of the effective source size and $\lambda$. The contribution of long-lived resonances ($c\tau > 5$~fm) decaying into charged pions and kaons, such as $\omega$, $\phi$, can be considered in the RSM as feed-down from weak decays. In this context, $12.2\%$ ($6\%$) of the charged pions (kaons) can be modeled as secondary particles from weak decays. To take this contribution into account, the $\lambda$ parameter related to the genuine part has to be corrected for the corresponding abundances of long-lived resonances. The fully corrected  $\lambda_{\mathrm{gen}}$ parameters in different \kt (\mt) intervals are presented in Table~\ref{tab:lambdaParam} for \spipi and \Kp pairs. An uncertainty of 10\% has been assigned to the data-driven evaluated values, accounting for possible inaccuracies of the determination of the fractions. 
The total uncertainty of the analysis is evaluated by employing a bootstrap procedure~\cite{Bootstrap}. This involves two main steps, firstly, randomly sampling from all of the aforementioned systematic variations, secondly, resampling the individual bins of the correlation function based on the statistical uncertainties. Subsequently, the fit procedure is repeated multiple times and a distribution of radii obtained. The central interval of the radii distribution, within which 68\% of the mass is concentrated, is calculated and determines the total uncertainty of the analysis. The systematic uncertainty has been isolated by repeating the procedure without resampling for statistical fluctuations. Notably, the dominating source of the systematic uncertainty is the variation of $\lambda$ parameters, contributing approximately 30\% to the total uncertainty.

\begin{table}[!ht]
\begin{center}
\caption{Compilation of $\lambda_{\mathrm{gen}}$ parameters for \kt (\mt) intervals. The values are derived from the fit results of the MC DCA templates to the experimental data, and from purity studies with Pythia~8.2 and GEANT~3 and they are corrected for the abundances of long-lived strongly decaying resonances.}
\begin{tabular}{|c| c | c | c | c | c |}
\hline\hline
(HM/MB) \kt (\GeVc) & 0.15--0.30 & 0.30--0.50 & 0.50--0.70 & 0.70--0.90 & 0.90--1.50 \\
\hline 
$\lambda_{\mathrm{gen}}$(\spipi) & 0.66~$\pm$~0.07 & 0.66~$\pm$~0.07 & 0.66~$\pm$~0.07 & 0.66~$\pm$~0.07 & 0.66~$\pm$~0.07\\
\hline\hline
(HM) \mt (\GeVmass) & 1.2--1.4 & 1.4--1.5 & 1.5--1.8 & 1.8--2.0 & -- \\
\hline 
$\lambda_{\mathrm{gen}}$(\Kp) & 0.76~$\pm$~0.08 & 0.73~$\pm$~0.08 & 0.66~$\pm$~0.07 & 0.65~$\pm$~0.07 & --\\
\hline \hline
\end{tabular}

\label{tab:lambdaParam}
\end{center}
\end{table}

%% file: Content/E_Results_and_discussion.tex
\section{Results and Discussion}\label{sec:results}
The same-sign \spipi correlation functions in MB collisions are measured in three multiplicity intervals and five different \kt ranges. 
As an example, correlation functions in the range of \kt (0.50$-$0.70) \GeVc and $N_{\mathrm{ch}}>~30$ are shown in
Fig.~\ref{fig:Ck_pipi_MB_example}. All correlations for the MB dataset are shown in the appendix in 
Figs.~\ref{fig:Ck_pipi_MB_mult0},~\ref{fig:Ck_pipi_MB_mult1}, and~\ref{fig:Ck_pipi_MB_mult2}. 
The systematic uncertainties are represented by boxes. The fits are performed for different \kt and multiplicity classes. 
On the left (right) panels, the results are shown assuming a polynomial of first (second) degree to describe the baseline of the \CF. 
The strength of the correlation function is sometimes underestimated in the fits with a linear baseline and better accommodated if the 
second degree polynomial is used to describe the residual background. The \spipi correlation lies above unity for low values of \ks, 
due to the enhancement stemming from Bose$-$Einstein quantum statistics, dominating the interaction. The repulsive Coulomb interaction
 has a visible impact on the correlation near $\ks = 0$ and forces \CF to deplete, where it finally overcomes 
 the Bose$-$Einstein quantum statistics. The same behavior is observed in the correlations of \spipi pairs obtained from the HM dataset. 
 An example of correlations for the \kt interval (0.70$-$0.90) \GeVc for HM events is shown in Fig.~\ref{fig:Ck_pipi_HM_example}. 
 All correlations of the HM dataset are shown in the appendix in Fig.~\ref{fig:Ck_pipi_HM}. 
 As for the MB results, the fits for the assumption of a polynomial of first (second) degree are shown on the left (right) panel. 
 With increasing \kt, the onset of the \CF shifts systematically to higher \ks. 
 This behavior is difficult to capture in the currently employed fitting procedure, and therefore may motivate the development
  of a more refined model for the correlation functions. However, it is also already known that in the case of \spipi correlations
   the fit quality improves if a 3D analysis is conducted~\cite{ALICE3Danapion, ALICEpipiInpp}. Nevertheless, the overall shapes
    of the \spipi correlations are captured by the fits in all studied \mt ranges. 

\begin{figure}[htb!]
\centering
\includegraphics[width=.47\textwidth]{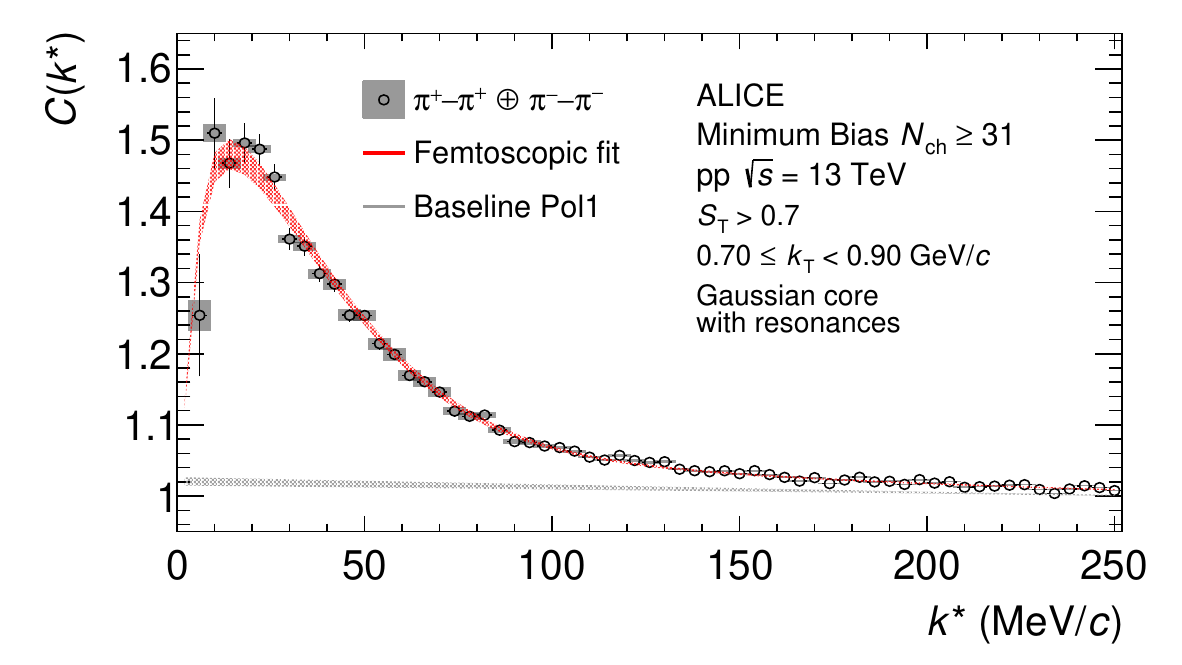}
\includegraphics[width=.47\textwidth]{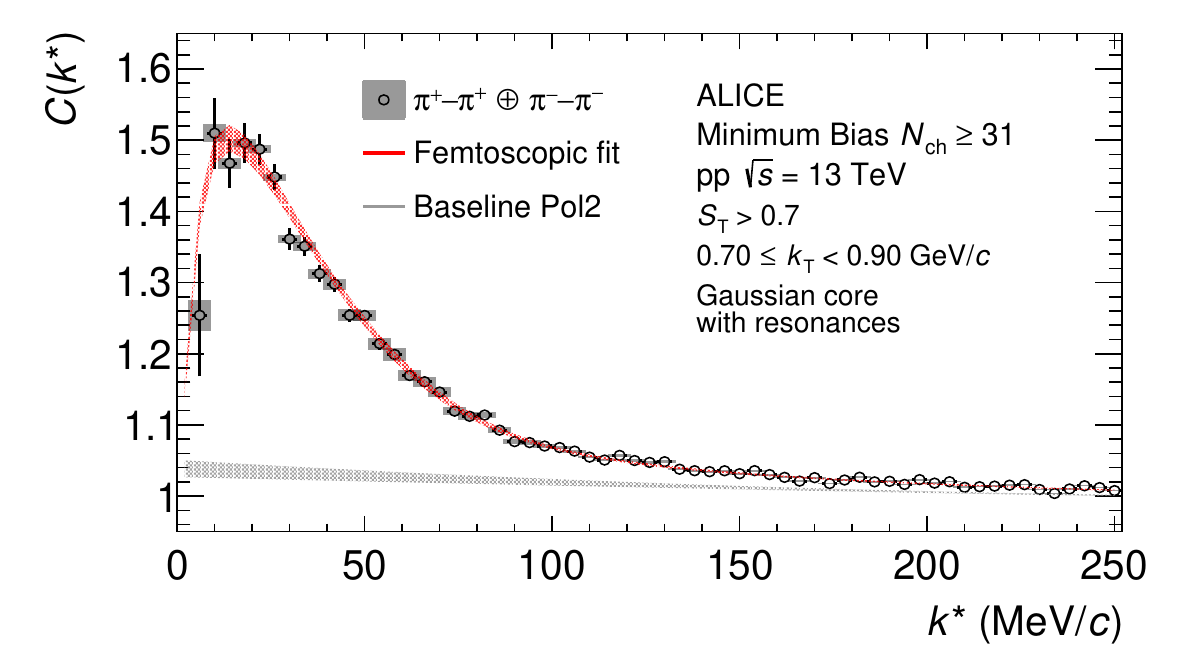}
\caption{Correlation function of \spipi pairs in the third (0.50--0.70 \GeVc) \kt interval for $N_{\mathrm{ch}}>~30$, the fits are performed using CATS~\cite{CATS} employing the RSM. The left (right) panel shows the results assuming a polynomial of first (second) degree as background. The statistical and systematic uncertainties are represented by the bars and rectangles, respectively. The uncertainty bands of the fit function are obtained by employing a bootstrap~\cite{Bootstrap} procedure.}
\label{fig:Ck_pipi_MB_example}
\end{figure}

\begin{figure}[htb!]
\centering
\includegraphics[width=.47\textwidth]{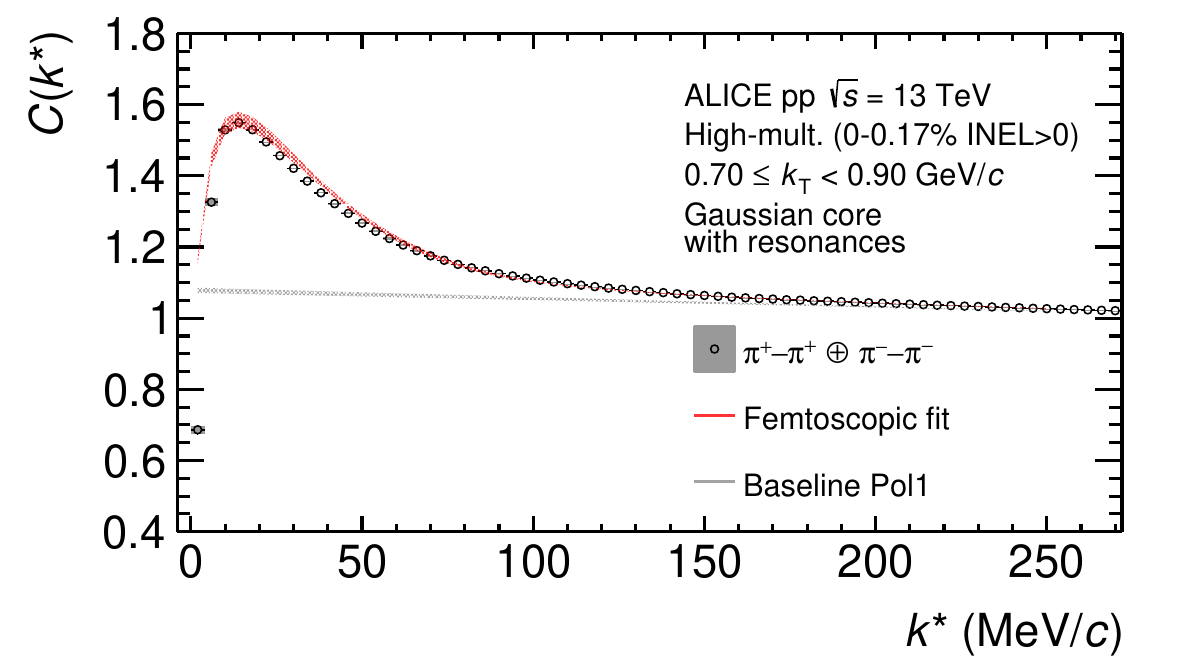}
\includegraphics[width=.47\textwidth]{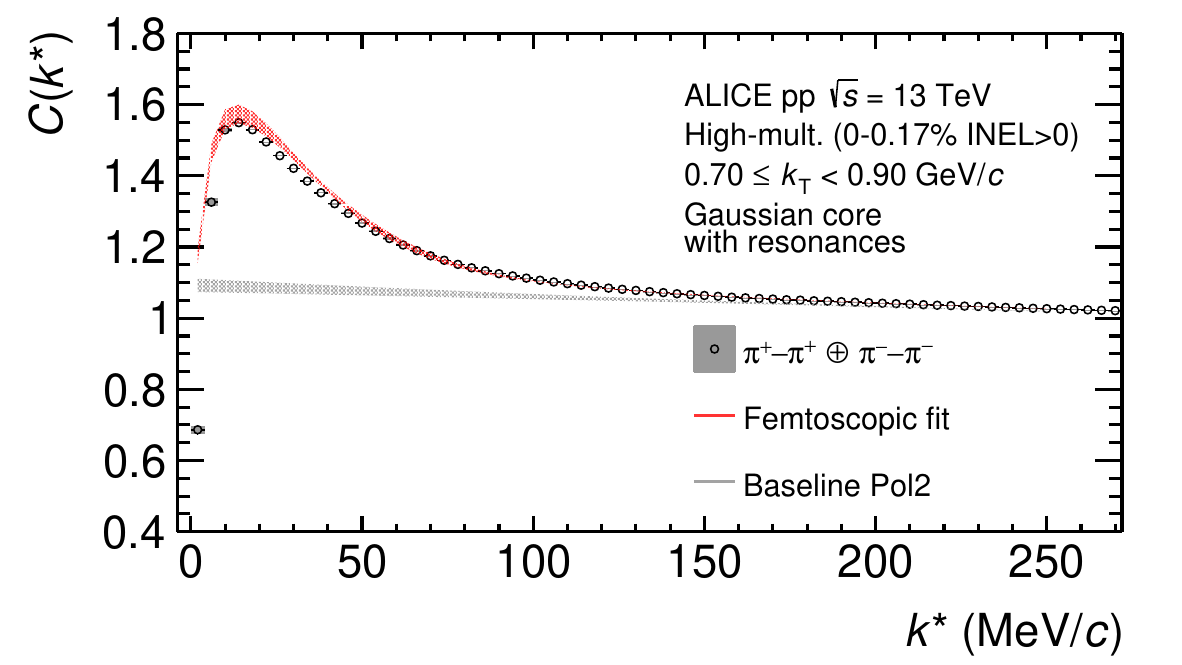}
\caption{Correlation function of \spipi pairs in the fourth (0.70--0.90 \GeVc) \kt interval for HM collisions, the fits on the left (right) are performed using CATS~\cite{CATS} employing the RSM and assuming a polynomial of first (second) degree as background. The statistical and systematic uncertainties are represented by the bars and rectangles, respectively. The uncertainty bands of the fit function are obtained by employing a bootstrap~\cite{Bootstrap} procedure.}
\label{fig:Ck_pipi_HM_example}
\end{figure}

\begin{figure}[htpb!]
\centering
\includegraphics[width=.45\textwidth]{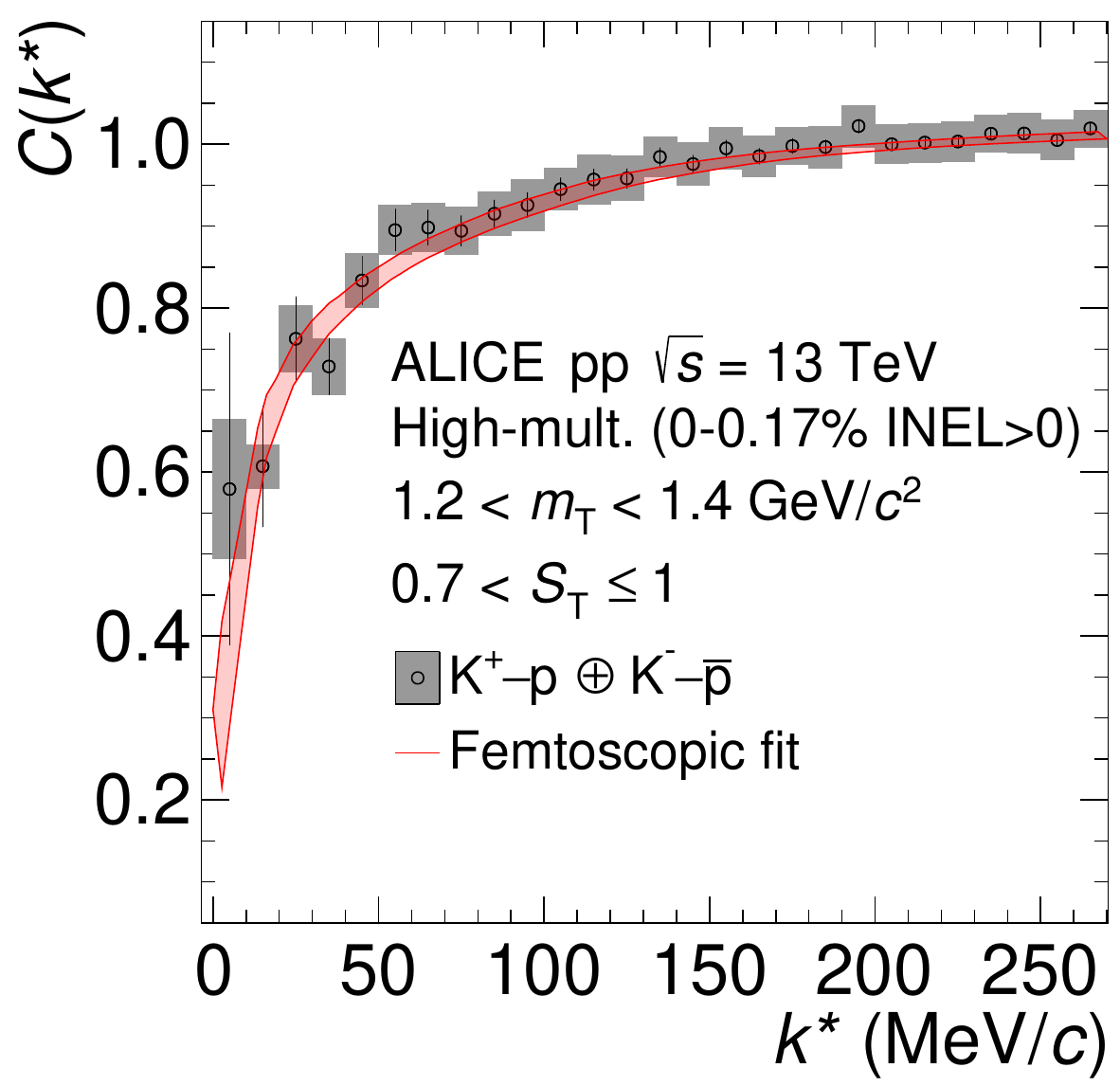}
\includegraphics[width=.45\textwidth]
{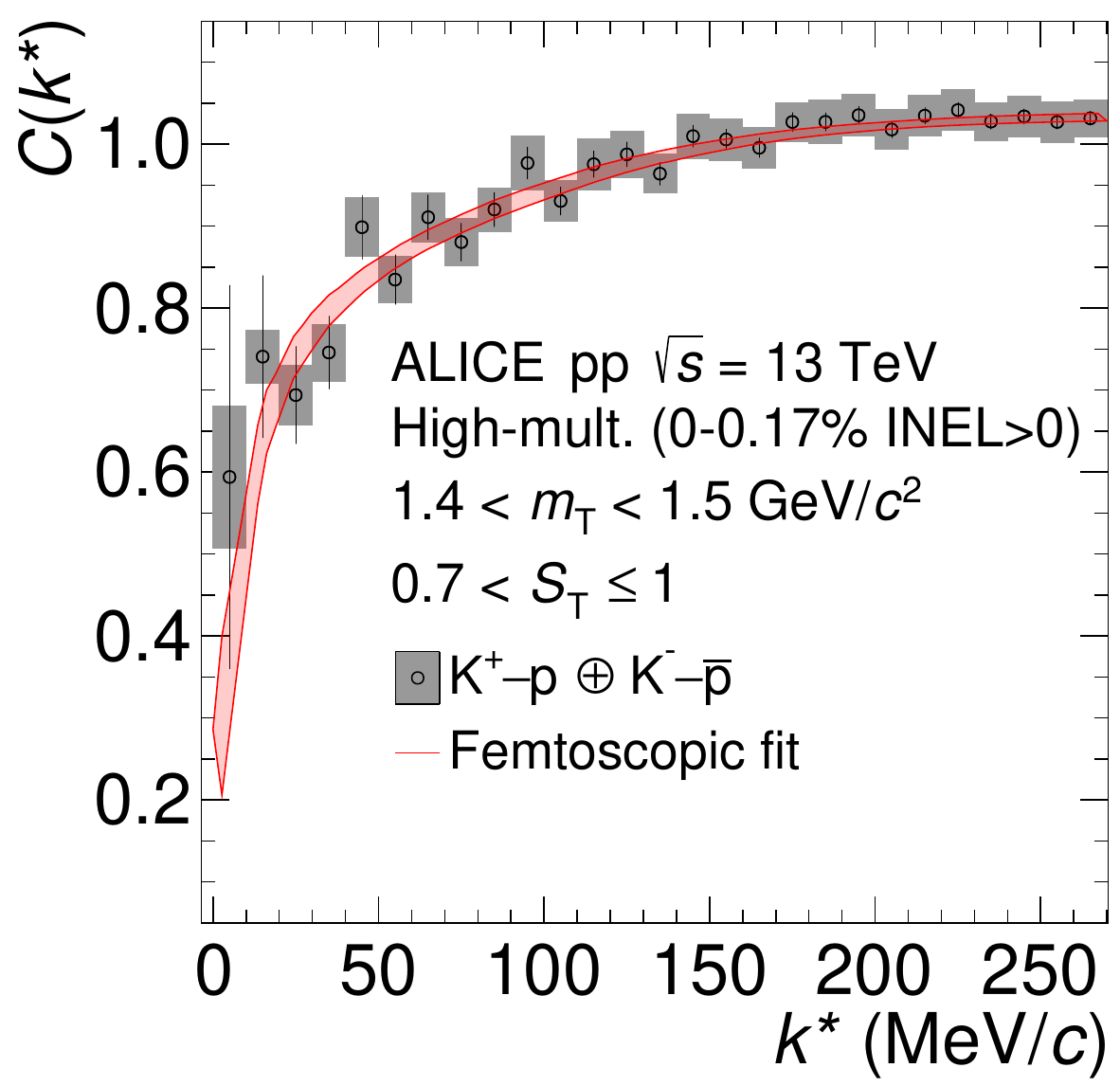}
\\
\includegraphics[width=.45\textwidth]{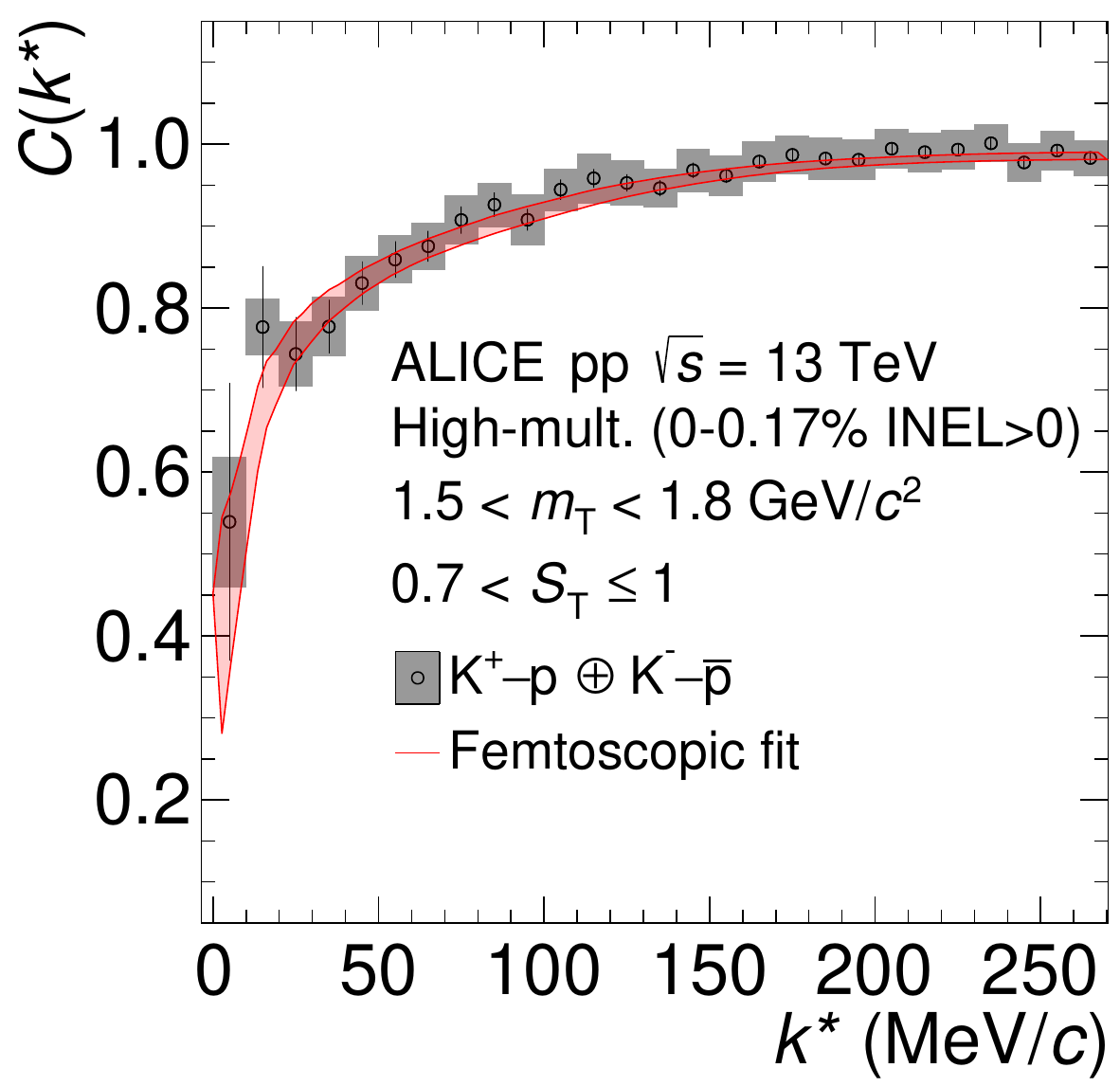}
\includegraphics[width=.45\textwidth]{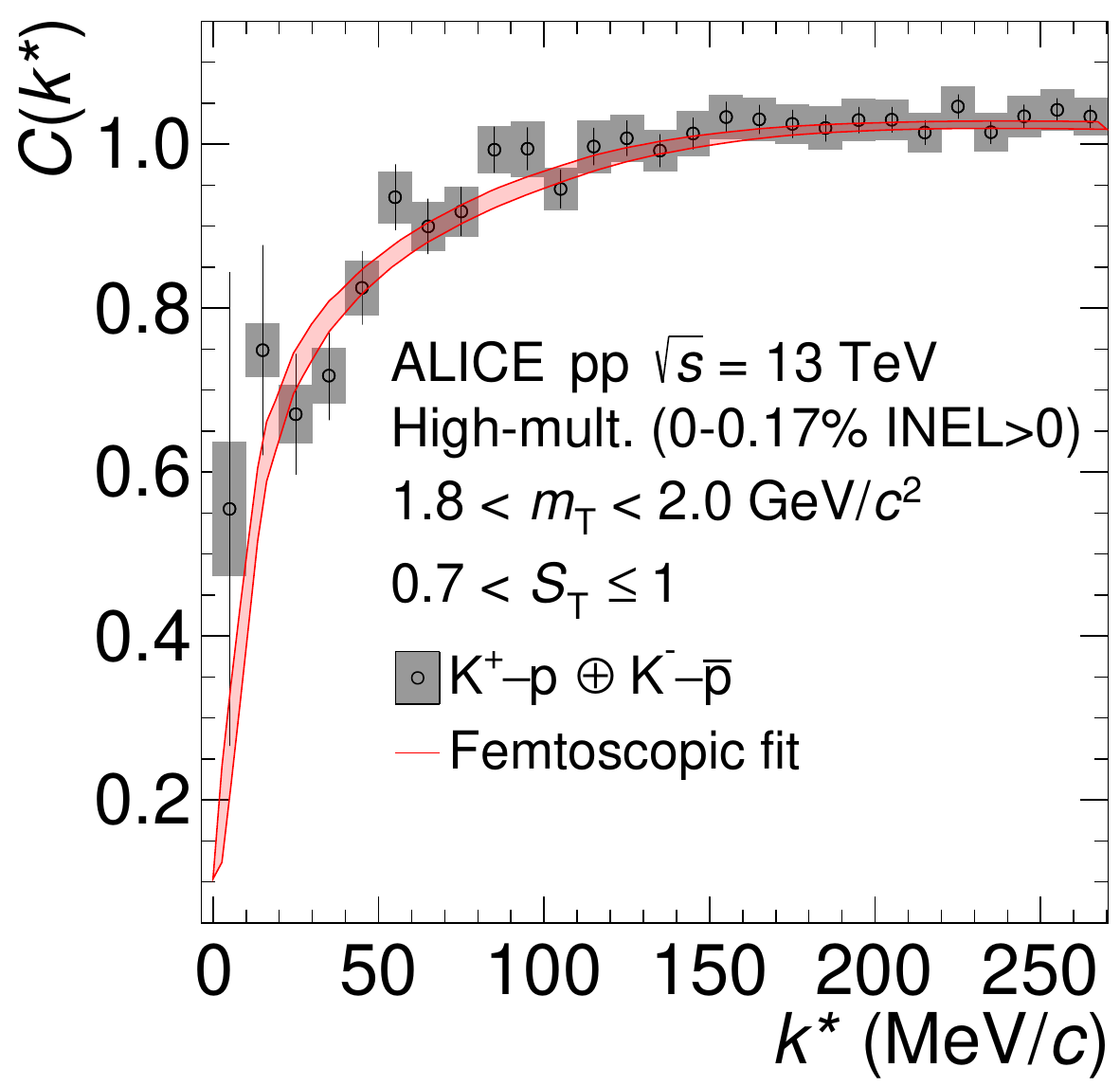}
\caption{Correlation functions of \Kp pairs in different ranges of \mt for HM pp collisions as indicated in each panel. Vertical bars represent statistical uncertainties while boxes systematic uncertainties, respectively. The fits are performed using CATS~\cite{CATS} employing the root mean square deviation for a uniform distribution in each \ks interval. The uncertainty bands of the fit function are obtained by employing a bootstrap~\cite{Bootstrap} procedure.}
\label{fig:Ck_Kp_HM}
\end{figure}

The resulting \CF for same-sign \Kp pairs in HM events are shown in Fig.~\ref{fig:Ck_Kp_HM}. The correlation function lies below unity for low values of \ks indicating an overall repulsive interaction. The specific shape of the \Kp correlation is the result of the interplay between the repulsive Coulomb and strong interactions. 

\begin{figure}[hbt]
\centering
\includegraphics[width=.80\textwidth]{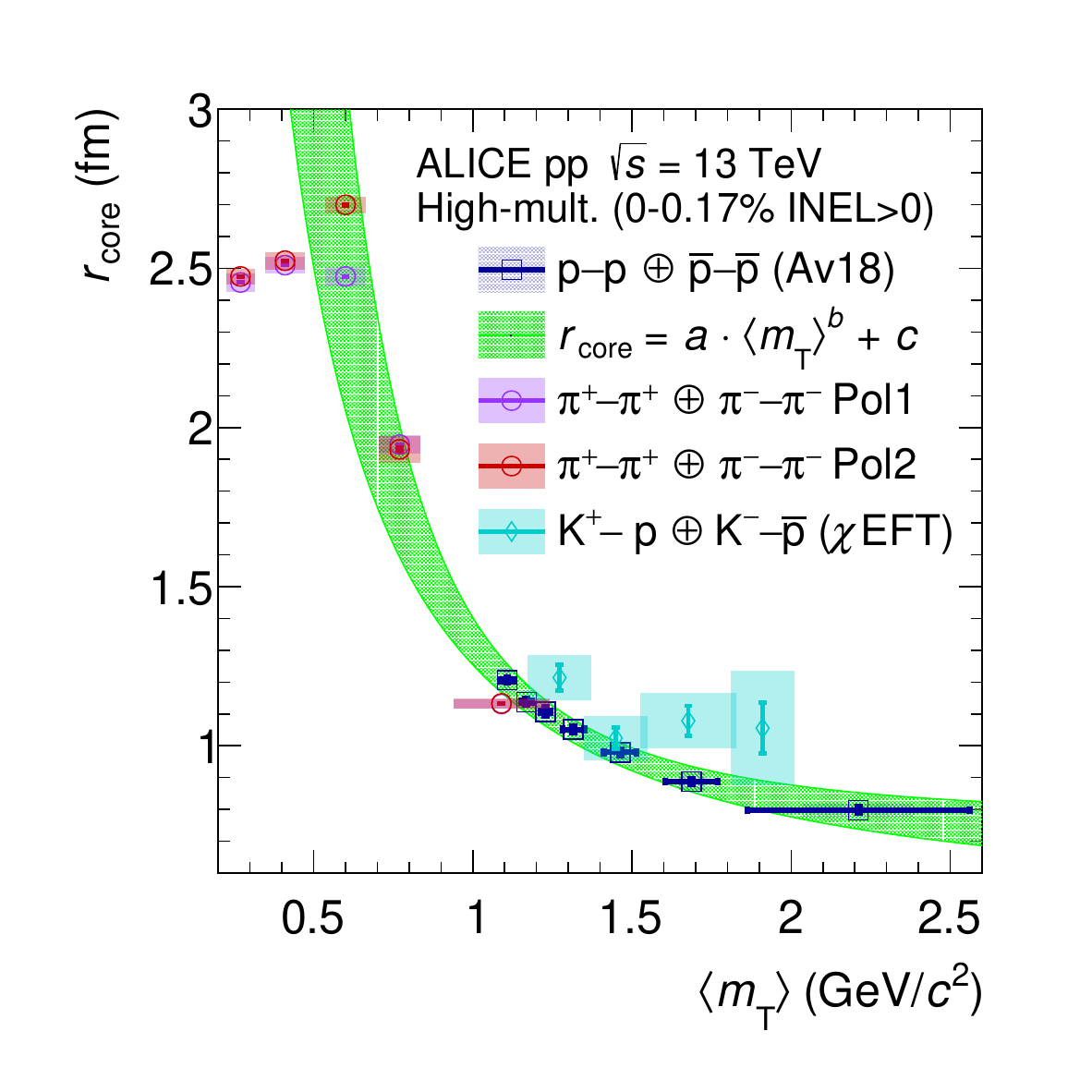}
\caption{Extracted radii from the fit as a function of \mt for the HM analysis of meson$-$meson, meson$-$baryon~(this work), and baryon$-$baryon~\cite{CommonSource} correlations. The green band corresponds to the parametrization of the \mt scaling of the \pP correlations and is shown with the associated $3\sigma$ spread.}
\label{fig:RadScaling_HM_BB_comb}
\end{figure}

\begin{figure}[hbt]
\centering
\includegraphics[width=1.\textwidth]{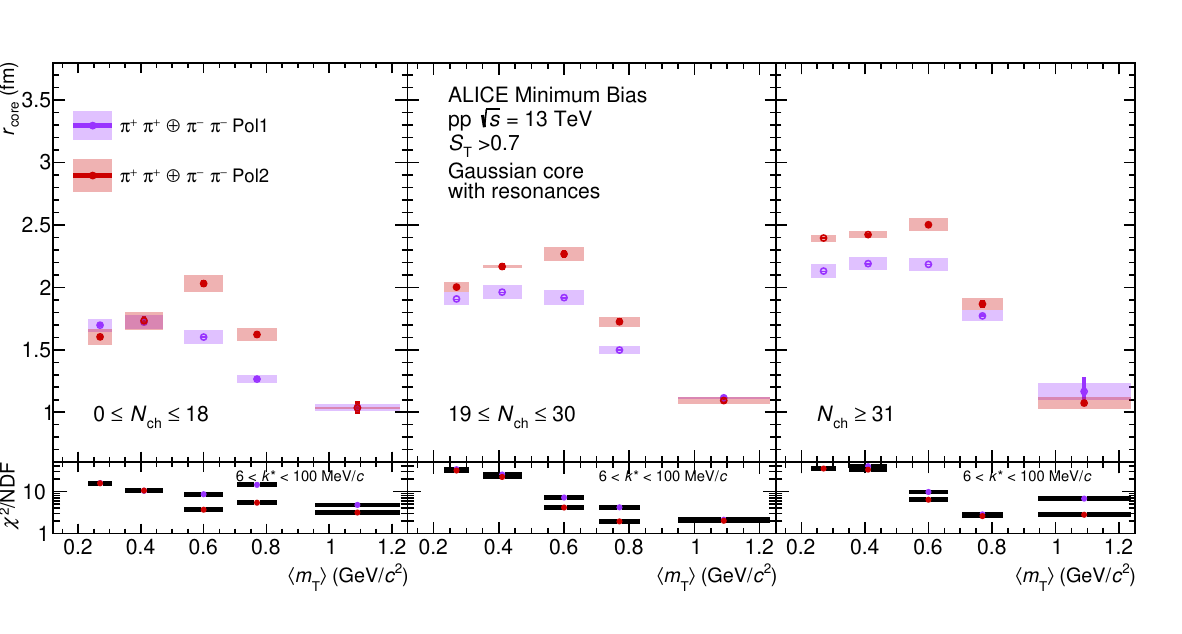}
\caption{Extracted radii from the fits of \spipi correlations as a function of \mt for three different multiplicity classes. From left to right and bottom: $1 \leq N_{\mathrm{ch}} \leq 18$, $19 \leq N_{\mathrm{ch}} \leq 30$ and $N_{\mathrm{ch}} \geq 31$. The reduced $\chi^2$ for the obtained radii is shown in the lower panels and evaluated in the range of $6< \ks <100$~\MeVc. The assumption that a second-order polynomial background shape exhibits either greater or equivalent compatibility with the data compared to a first-order polynomial.} 
\label{fig:RadScaling_MB}
\end{figure}

In Figure~\ref{fig:RadScaling_HM_BB_comb}, the results of the Gaussian core radius \rc for \spipi and \Kp pairs in HM pp collisions at \thirteen are shown as a function of \mt together with the radii obtained from \pP correlations~\cite{CommonSource}. The agreement of the meson$-$meson and meson$-$baryon \rc with the \mt scaling (green band) of the \pP correlations is remarkable and provides additional support for the scenario of a common emitting source for all hadrons in small systems. Furthermore, the results presented in Fig.~\ref{fig:RadScaling_MB} for the three multiplicity intervals considered for MB pp collisions at \thirteen exhibit the intuitive behavior of growing \rc with increasing multiplicity and also show the scaling of \rc with increasing \mt. The degree of consistency between the data and the model is estimated by using the $\chi^{2} = \sum_i (x_{\mathrm{Data}}^i - x_{\mathrm{Theory}}^i)^2/(\sigma_{\mathrm{Data}}^i)^2$ normalized to the number of degrees of freedom (NDF), commonly called the reduced $\chi^{2}$. The bottom panels in Fig.~\ref{fig:RadScaling_MB} show the reduced $\chi^{2}$ for $6< \ks< 100$~\MeVc, indicating that the assumption of a polynomial of second degree always performs as good or better than the first degree choice. In all the multiplicity intervals, for \mt below 0.6~\GeVmass, which is probed with the \spipi correlations, the \rc reaches a ``saturation'' regime and becomes independent of \mt.  For \Kp pairs no breaking of the \mt scaling is observed, however, it should be noted that the lowest \mt range accessed for \Kp correlations in this analysis is 1.2--1.4 \GeVmass while the theoretical lower limit is at $0.7$ \GeVmass, leaving some narrow phase-space for a possible violation of the \mt scaling. A similar change in trend was already observed by CMS~\cite{CMSpipiInpp} and ALICE~\cite{ALICEpipiInpp} in pp collisions at \thirteen and 900~\gev. A possible explanation for the observed trend could be related to the fact that for low-\mt pairs in \pp collisions, the region of homogeneity~(\rc) spans the entire physical extension of the hadronization hypersurface which, hence, represents the observed upper boundary of \rc. Any more detailed explanation will most likely require a consistent hydrodynamical treatment of small systems and a careful study of the properties of the resulting hadronization hypersurface and, hence, is outside the scope of this paper. However, these data provide valuable input for transport models which should describe the spatial extension of the particle emitting source in small systems and may inspire new theoretical works towards the description of the observed \rc scaling with \mt for all the measured particle pairs.

%% file: Content/F_Summary.tex
\section{Summary}
\label{sec:conclusion}
In this work, the spatial extension of the particle source function is studied using correlations of the relative momentum of particles, referred to as femtoscopy. The source function is described by a Gaussian core and an exponential resonance halo~\cite{CommonSource}. The resonance abundances are fixed from estimations obtained within the canonical statistical hadronization model and the specific decay kinematics from the EPOS~\cite{EPOS} event generator. This prescription allows for the first time a quantitative description of the exponential type source for \spipi correlation functions in intervals of \mt. The correlation functions are studied for high multiplicity pp collisions and in multiplicity intervals for minimum bias collisions. The model is able to capture the widths of the studied correlation functions, however, cannot provide a full description of the correlation strength observed in the data for all studied multiplicity intervals, \mt ranges, and particle pairings. Further refinements of the employed model include an extension to a 3D analysis and the application of recently developed particle emission models such as CECA~\cite{CECA}, which for now was only tested for baryon$-$baryon correlations. Finally, by also analyzing the \Kp correlation functions in intervals of \mt within the same framework, the Gaussian core radii \rc are obtained and their dependence on \mt is studied. A \mt scaling for \rc consistent with the one reported in~\cite{CommonSource}, obtained from the analysis of baryon$-$baryon correlations, is found for $\mt > 0.6$~\GeVmass and, hence, points to a common emitting source for all hadrons in small systems at the LHC. For $\mt < 0.6$~\GeVmass, which is accessible by the \spipi correlation functions a saturation of \rc is observed leading to a breaking of the \mt scaling. A possible explanation could be that \rc spans the entire physical extension of the hadronization hypersurface and hence can grow no further.

%% file: fa_2023-10-27_Opt_C.tex

The ALICE Collaboration would like to thank all its engineers and technicians for their invaluable contributions to the construction of the experiment and the CERN accelerator teams for the outstanding performance of the LHC complex.
The ALICE Collaboration gratefully acknowledges the resources and support provided by all Grid centres and the Worldwide LHC Computing Grid (WLCG) collaboration.
The ALICE Collaboration acknowledges the following funding agencies for their support in building and running the ALICE detector:
A. I. Alikhanyan National Science Laboratory (Yerevan Physics Institute) Foundation (ANSL), State Committee of Science and World Federation of Scientists (WFS), Armenia;
Austrian Academy of Sciences, Austrian Science Fund (FWF): [M 2467-N36] and Nationalstiftung f\"{u}r Forschung, Technologie und Entwicklung, Austria;
Ministry of Communications and High Technologies, National Nuclear Research Center, Azerbaijan;
Conselho Nacional de Desenvolvimento Cient\'{\i}fico e Tecnol\'{o}gico (CNPq), Financiadora de Estudos e Projetos (Finep), Funda\c{c}\~{a}o de Amparo \`{a} Pesquisa do Estado de S\~{a}o Paulo (FAPESP) and Universidade Federal do Rio Grande do Sul (UFRGS), Brazil;
Bulgarian Ministry of Education and Science, within the National Roadmap for Research Infrastructures 2020-2027 (object CERN), Bulgaria;
Ministry of Education of China (MOEC) , Ministry of Science \& Technology of China (MSTC) and National Natural Science Foundation of China (NSFC), China;
Ministry of Science and Education and Croatian Science Foundation, Croatia;
Centro de Aplicaciones Tecnol\'{o}gicas y Desarrollo Nuclear (CEADEN), Cubaenerg\'{\i}a, Cuba;
Ministry of Education, Youth and Sports of the Czech Republic, Czech Republic;
The Danish Council for Independent Research | Natural Sciences, the VILLUM FONDEN and Danish National Research Foundation (DNRF), Denmark;
Helsinki Institute of Physics (HIP), Finland;
Commissariat \`{a} l'Energie Atomique (CEA) and Institut National de Physique Nucl\'{e}aire et de Physique des Particules (IN2P3) and Centre National de la Recherche Scientifique (CNRS), France;
Bundesministerium f\"{u}r Bildung und Forschung (BMBF) and GSI Helmholtzzentrum f\"{u}r Schwerionenforschung GmbH, Germany;
General Secretariat for Research and Technology, Ministry of Education, Research and Religions, Greece;
National Research, Development and Innovation Office, Hungary;
Department of Atomic Energy Government of India (DAE), Department of Science and Technology, Government of India (DST), University Grants Commission, Government of India (UGC) and Council of Scientific and Industrial Research (CSIR), India;
National Research and Innovation Agency - BRIN, Indonesia;
Istituto Nazionale di Fisica Nucleare (INFN), Italy;
Japanese Ministry of Education, Culture, Sports, Science and Technology (MEXT) and Japan Society for the Promotion of Science (JSPS) KAKENHI, Japan;
Consejo Nacional de Ciencia (CONACYT) y Tecnolog\'{i}a, through Fondo de Cooperaci\'{o}n Internacional en Ciencia y Tecnolog\'{i}a (FONCICYT) and Direcci\'{o}n General de Asuntos del Personal Academico (DGAPA), Mexico;
Nederlandse Organisatie voor Wetenschappelijk Onderzoek (NWO), Netherlands;
The Research Council of Norway, Norway;
Commission on Science and Technology for Sustainable Development in the South (COMSATS), Pakistan;
Pontificia Universidad Cat\'{o}lica del Per\'{u}, Peru;
Ministry of Education and Science, National Science Centre and WUT ID-UB, Poland;
Korea Institute of Science and Technology Information and National Research Foundation of Korea (NRF), Republic of Korea;
Ministry of Education and Scientific Research, Institute of Atomic Physics, Ministry of Research and Innovation and Institute of Atomic Physics and Universitatea Nationala de Stiinta si Tehnologie Politehnica Bucuresti, Romania;
Ministry of Education, Science, Research and Sport of the Slovak Republic, Slovakia;
National Research Foundation of South Africa, South Africa;
Swedish Research Council (VR) and Knut \& Alice Wallenberg Foundation (KAW), Sweden;
European Organization for Nuclear Research, Switzerland;
Suranaree University of Technology (SUT), National Science and Technology Development Agency (NSTDA) and National Science, Research and Innovation Fund (NSRF via PMU-B B05F650021), Thailand;
Turkish Energy, Nuclear and Mineral Research Agency (TENMAK), Turkey;
National Academy of  Sciences of Ukraine, Ukraine;
Science and Technology Facilities Council (STFC), United Kingdom;
National Science Foundation of the United States of America (NSF) and United States Department of Energy, Office of Nuclear Physics (DOE NP), United States of America.
In addition, individual groups or members have received support from:
Czech Science Foundation (grant no. 23-07499S), Czech Republic;
European Research Council, Strong 2020 - Horizon 2020 (grant nos. 950692, 824093), European Union;
ICSC - Centro Nazionale di Ricerca in High Performance Computing, Big Data and Quantum Computing, European Union - NextGenerationEU;
Academy of Finland (Center of Excellence in Quark Matter) (grant nos. 346327, 346328), Finland.

%% file: Content/Z_Appendix.tex
\section{Same Charge Pion Correlation Functions}
\label{Appendix:ppFits}

\begin{figure}[htbp!]
\centering
\includegraphics[width=.47\textwidth]{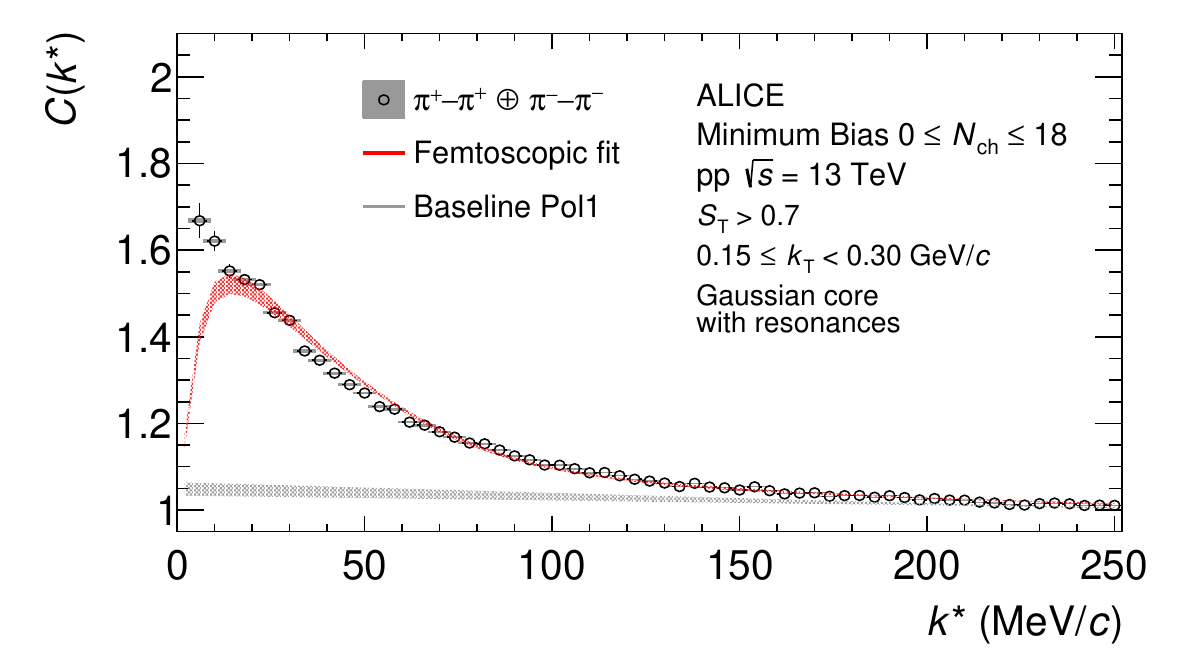}
\includegraphics[width=.47\textwidth]{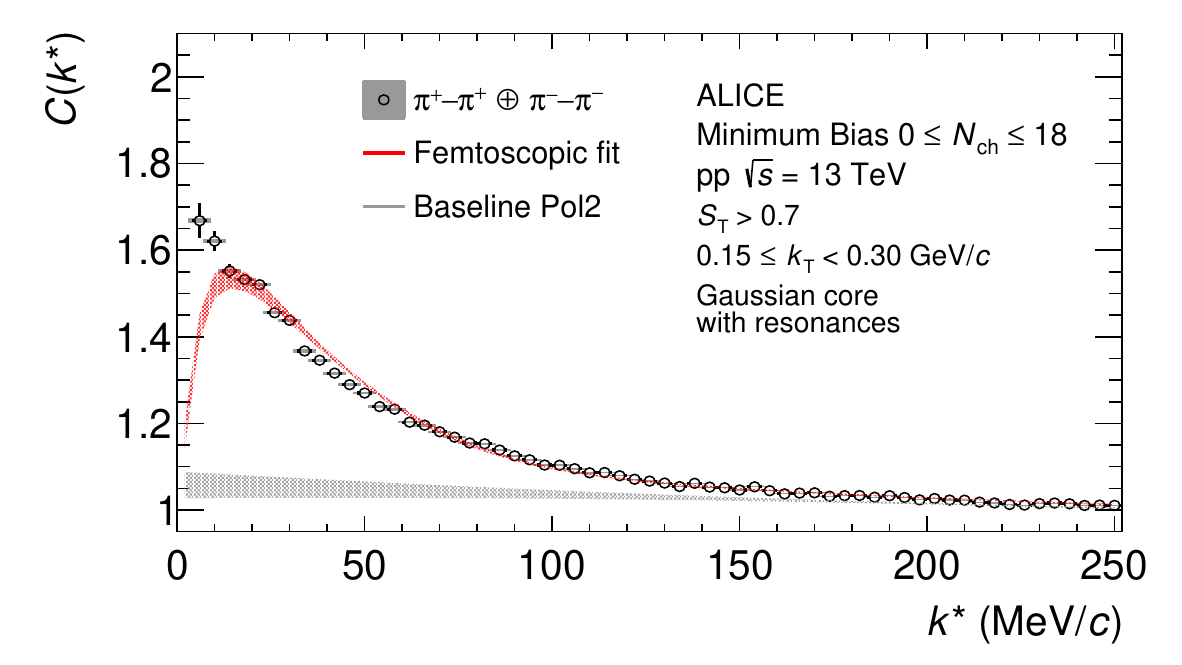}
\includegraphics[width=.47\textwidth]{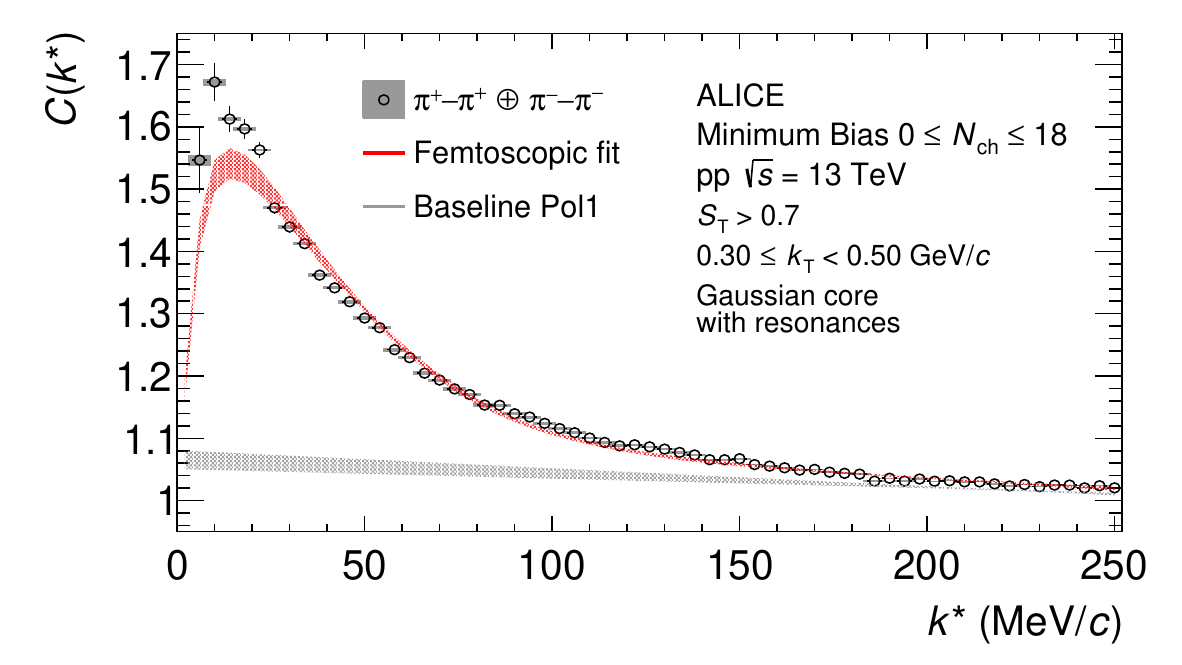}
\includegraphics[width=.47\textwidth]{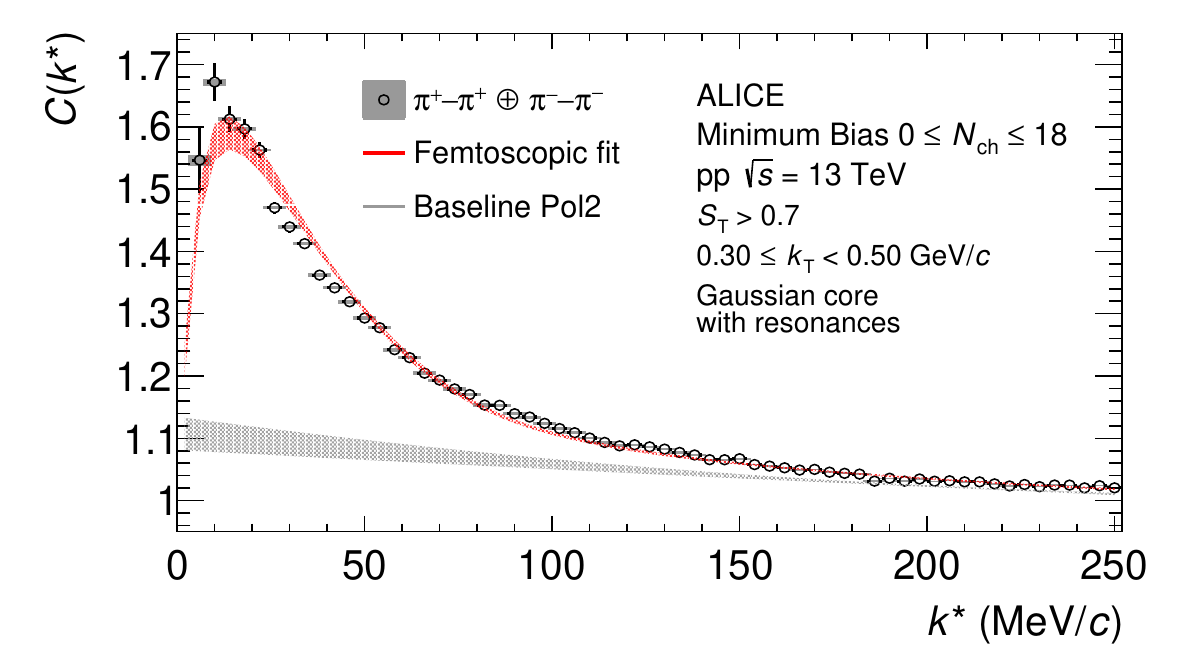}
\includegraphics[width=.47\textwidth]{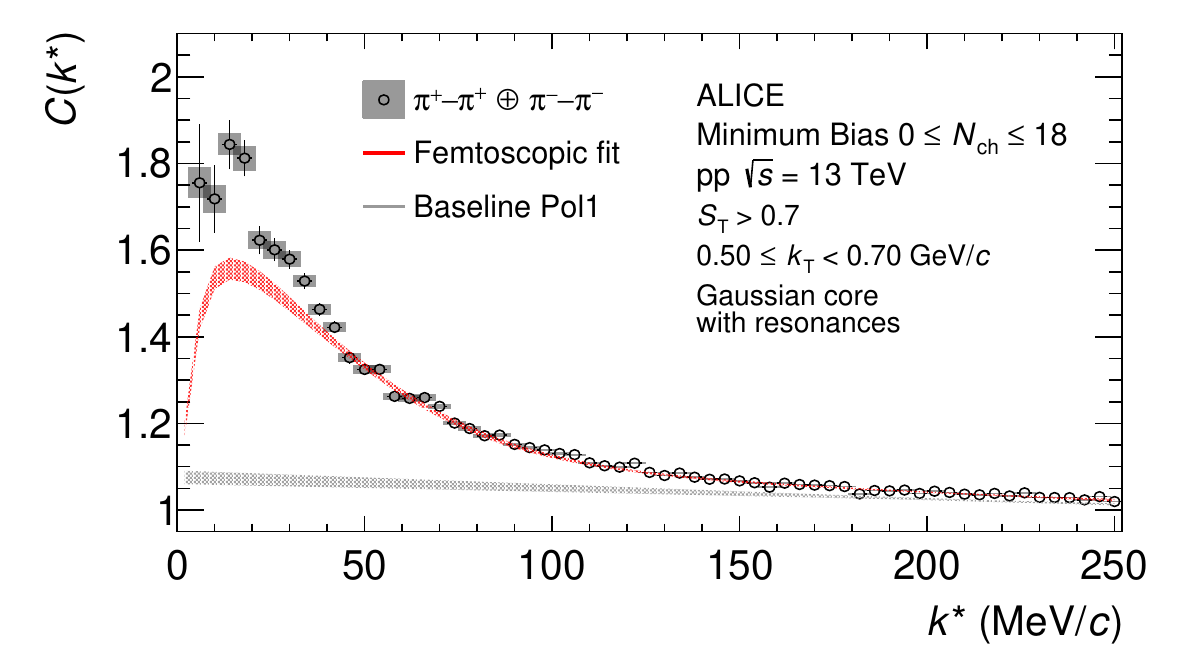}
\includegraphics[width=.47\textwidth]{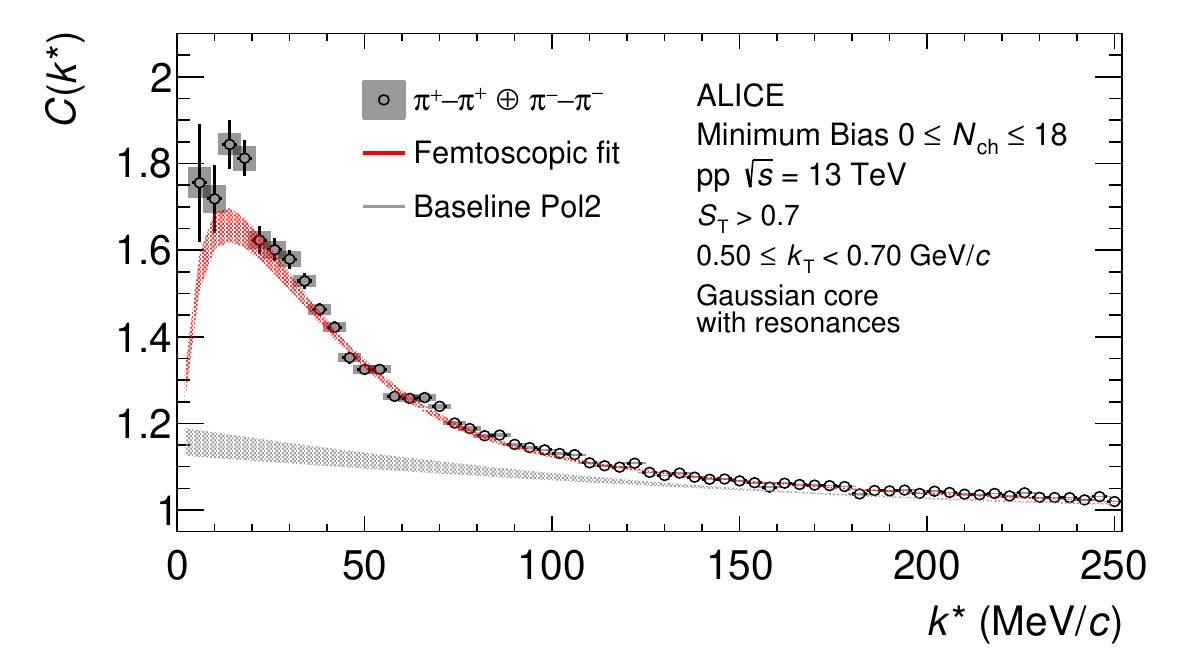}
\includegraphics[width=.47\textwidth]{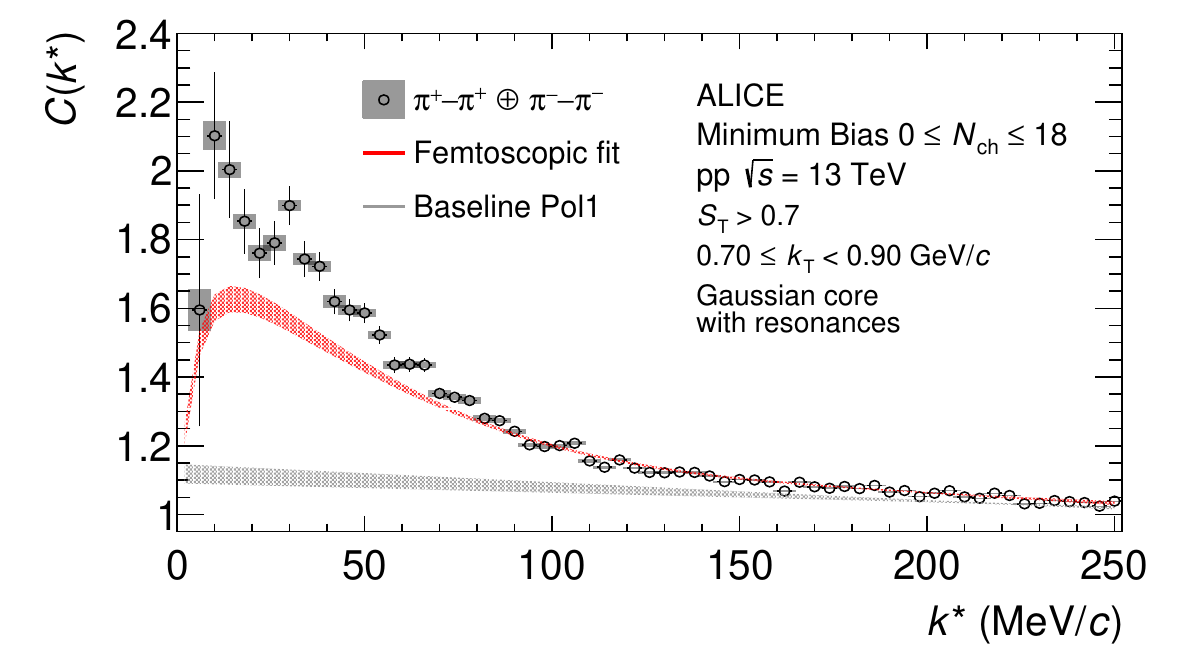}
\includegraphics[width=.47\textwidth]{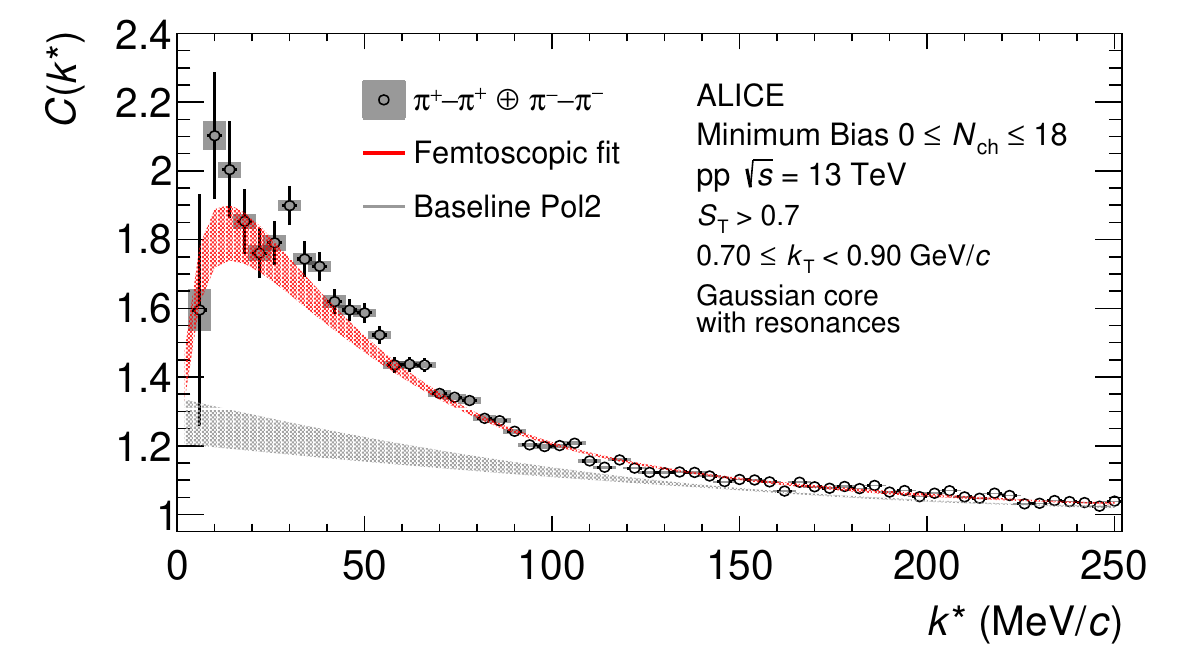}
\includegraphics[width=.47\textwidth]{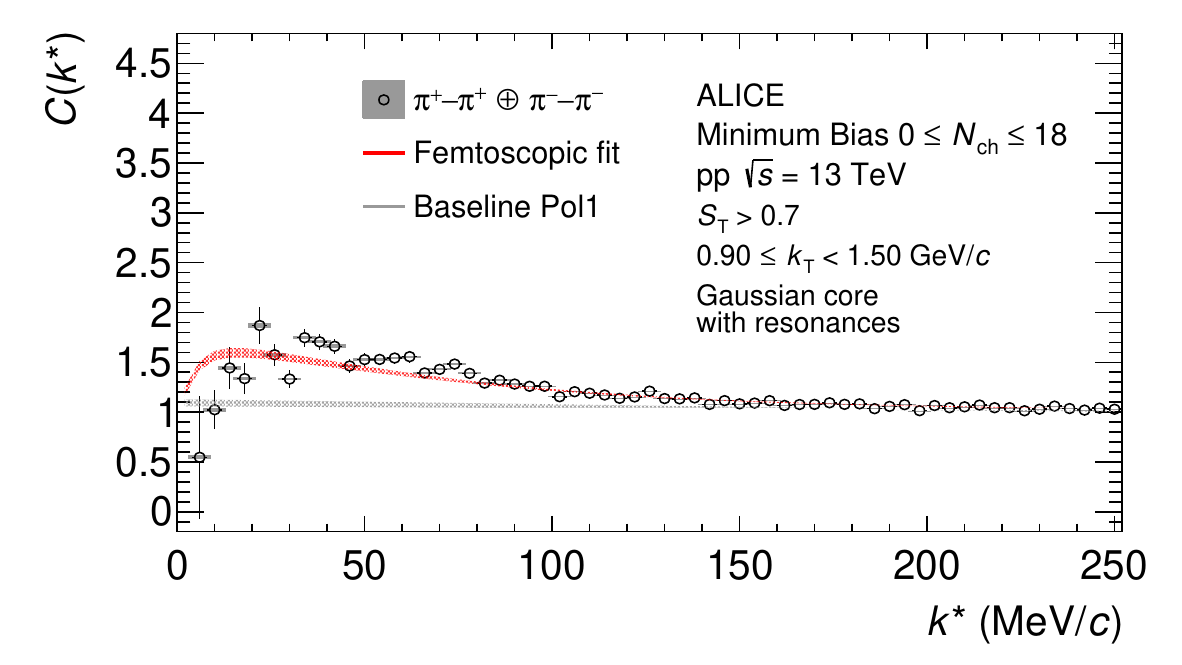}
\includegraphics[width=.47\textwidth]{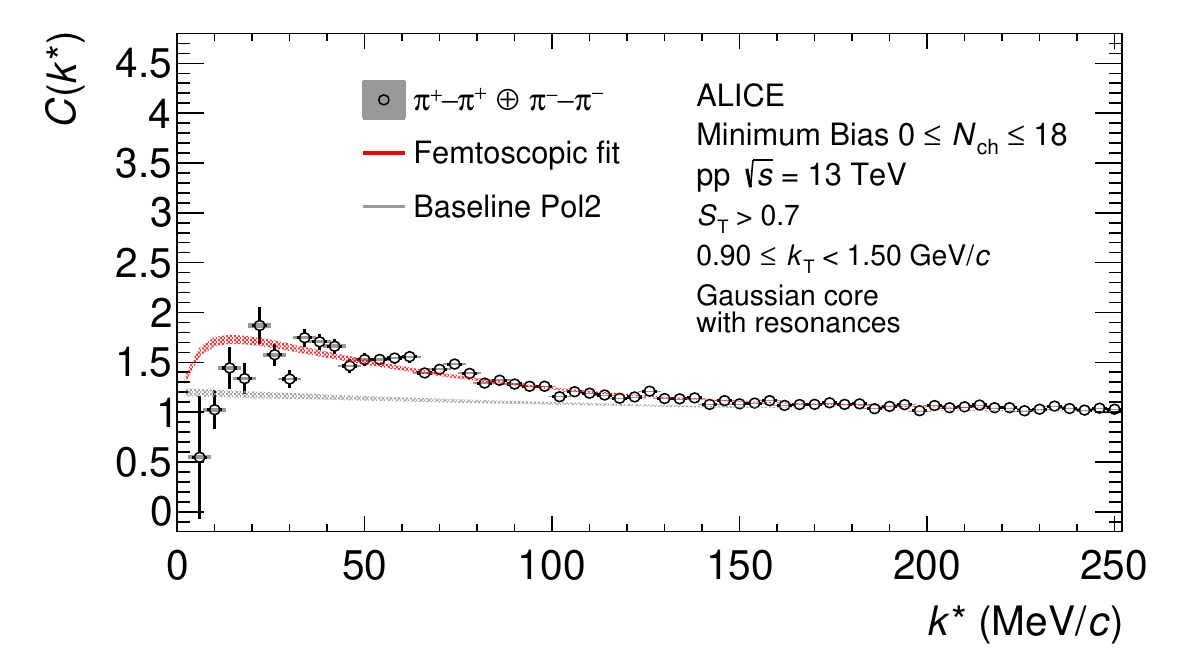}
\caption{Correlation functions of \spipi pairs in bins of \kt for $1\leq N_{\mathrm{ch}} \leq 18$, the fits are performed using CATS~\cite{CATS} employing the RSM. The left (right) panels show the results assuming a polynomial of first (second) degree as background. The uncertainties of the data points are obtained by varying the selection criteria as described in the text. The uncertainty bands of the fit function are obtained by employing a bootstrap~\cite{Bootstrap} procedure.}
\label{fig:Ck_pipi_MB_mult0}
\end{figure}

\begin{figure}[htbp!]
\centering
\includegraphics[width=.47\textwidth]{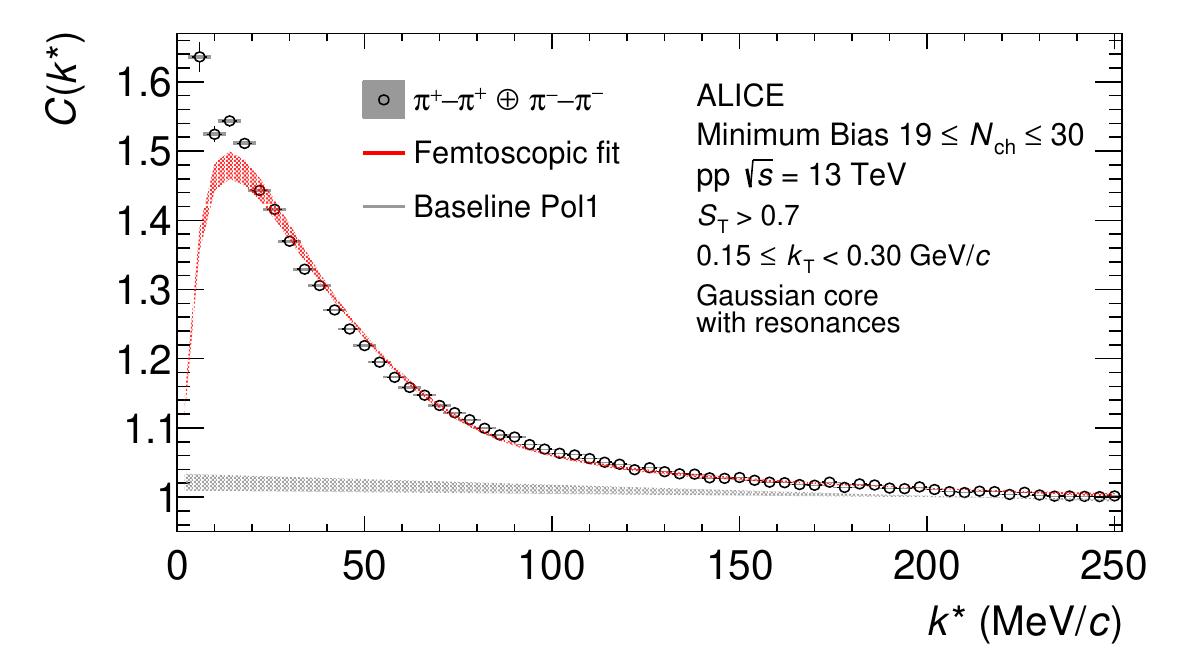}
\includegraphics[width=.47\textwidth]{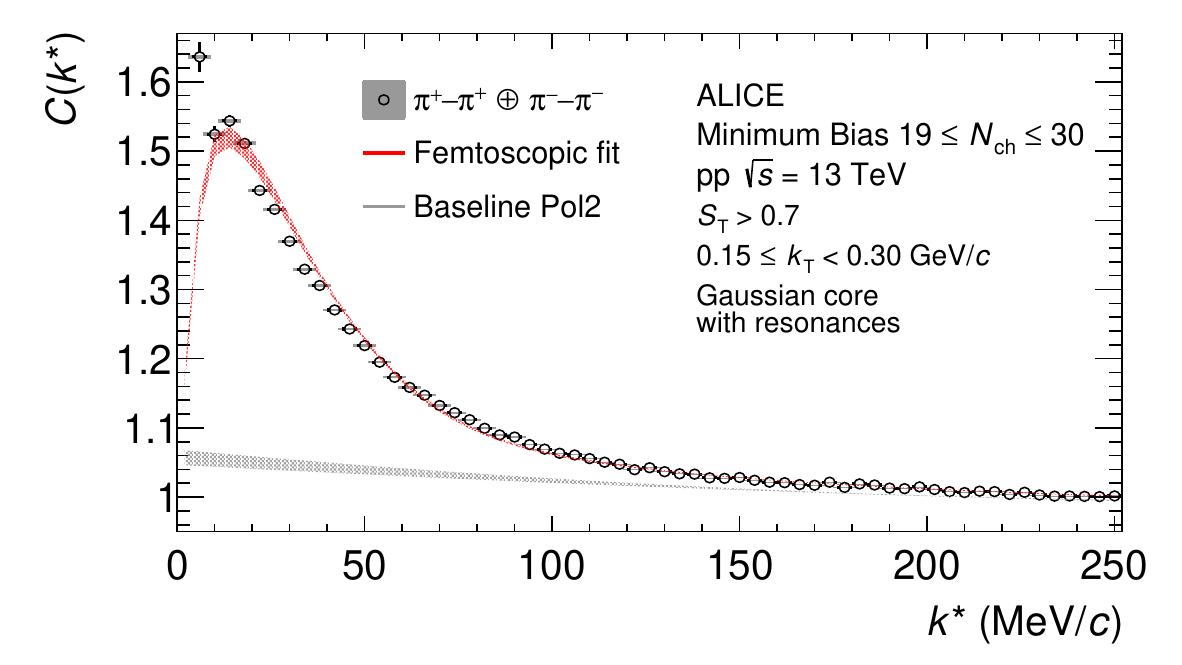}
\includegraphics[width=.47\textwidth]{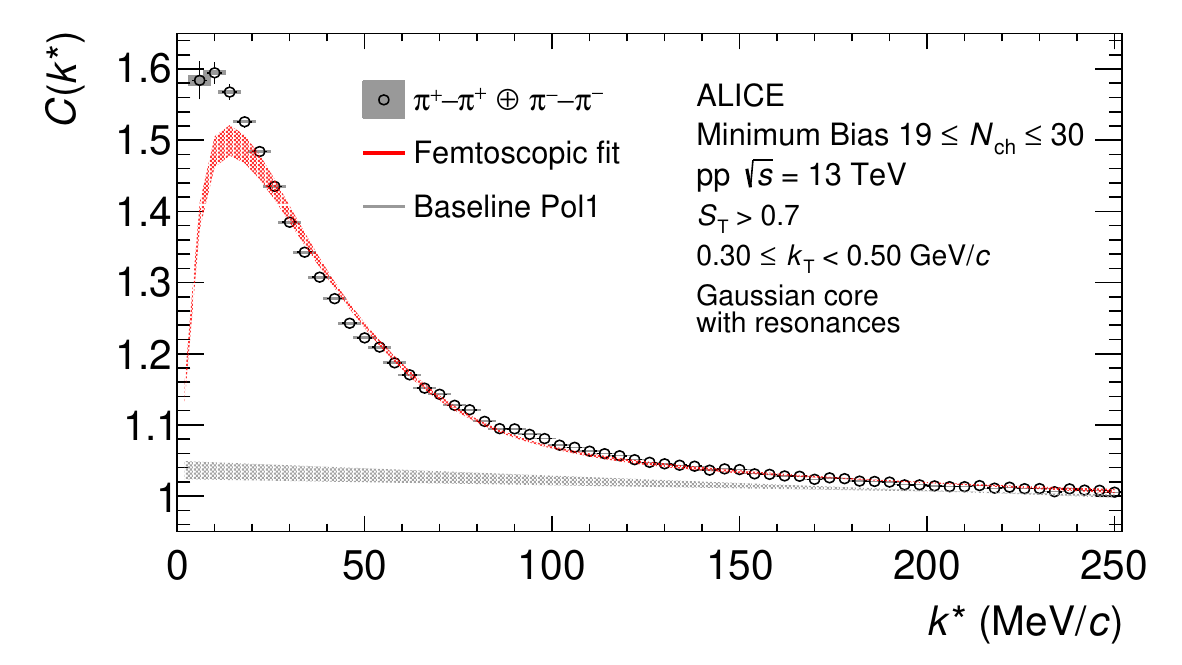}
\includegraphics[width=.47\textwidth]{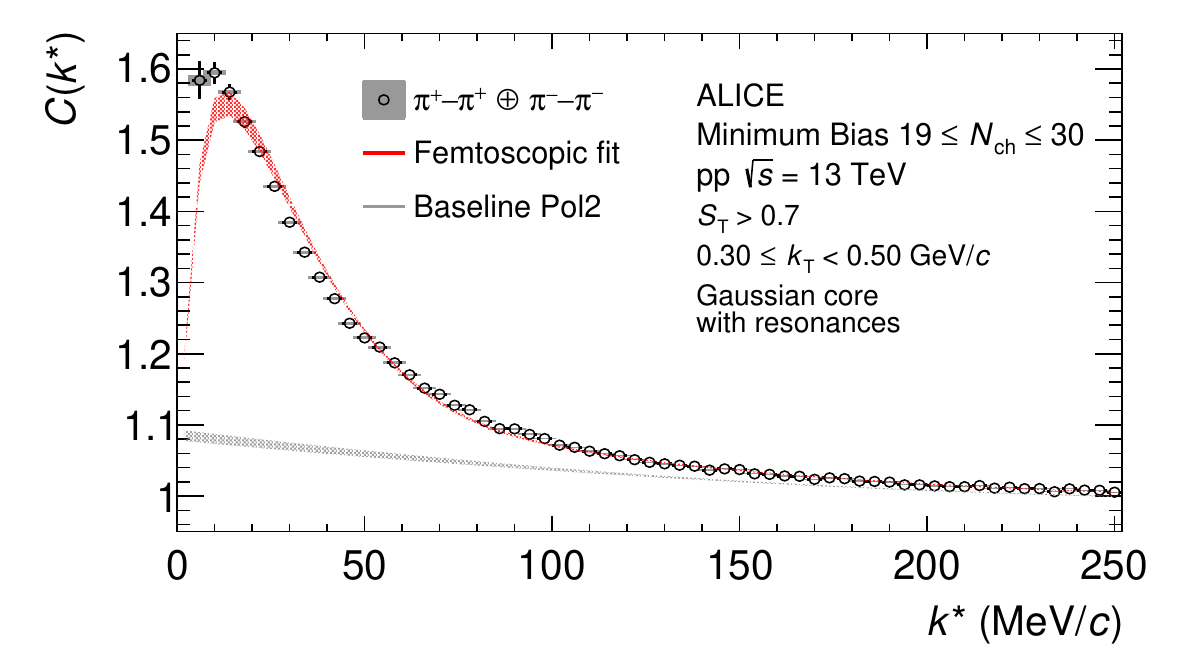}
\includegraphics[width=.47\textwidth]{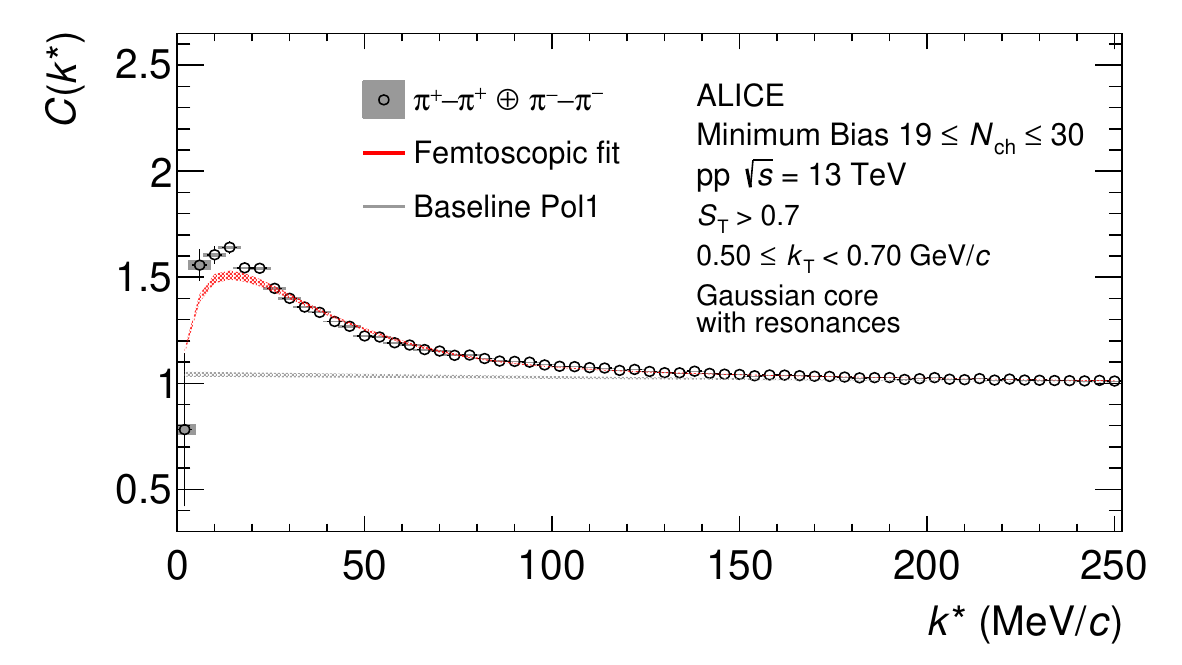}
\includegraphics[width=.47\textwidth]{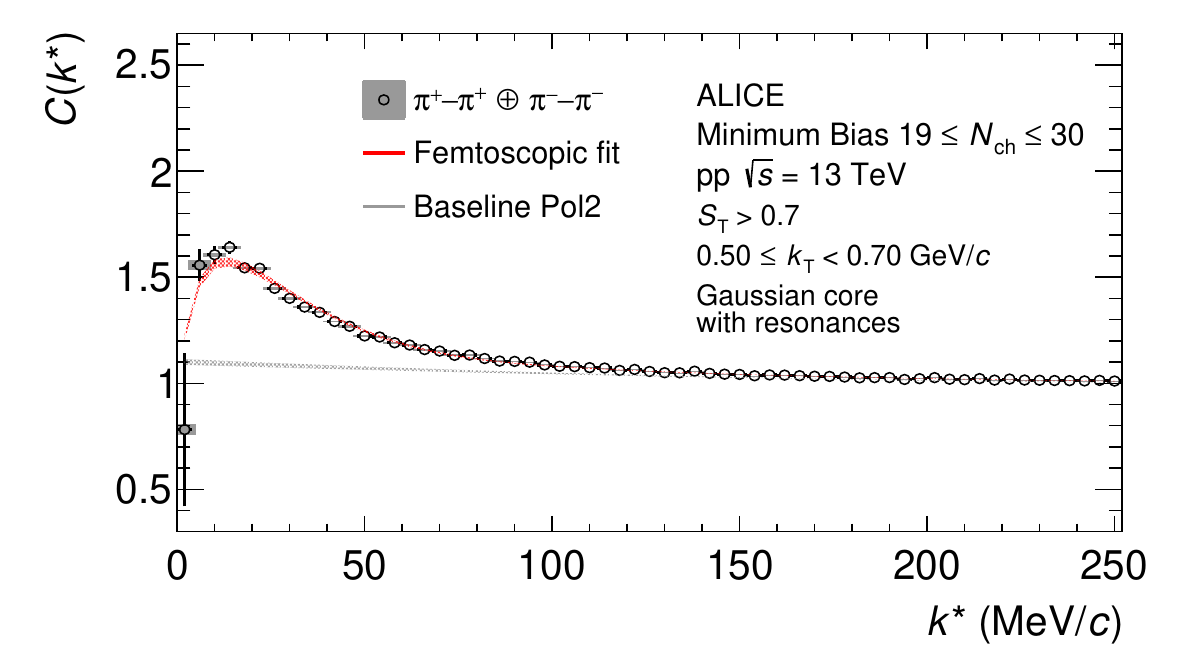}
\includegraphics[width=.47\textwidth]{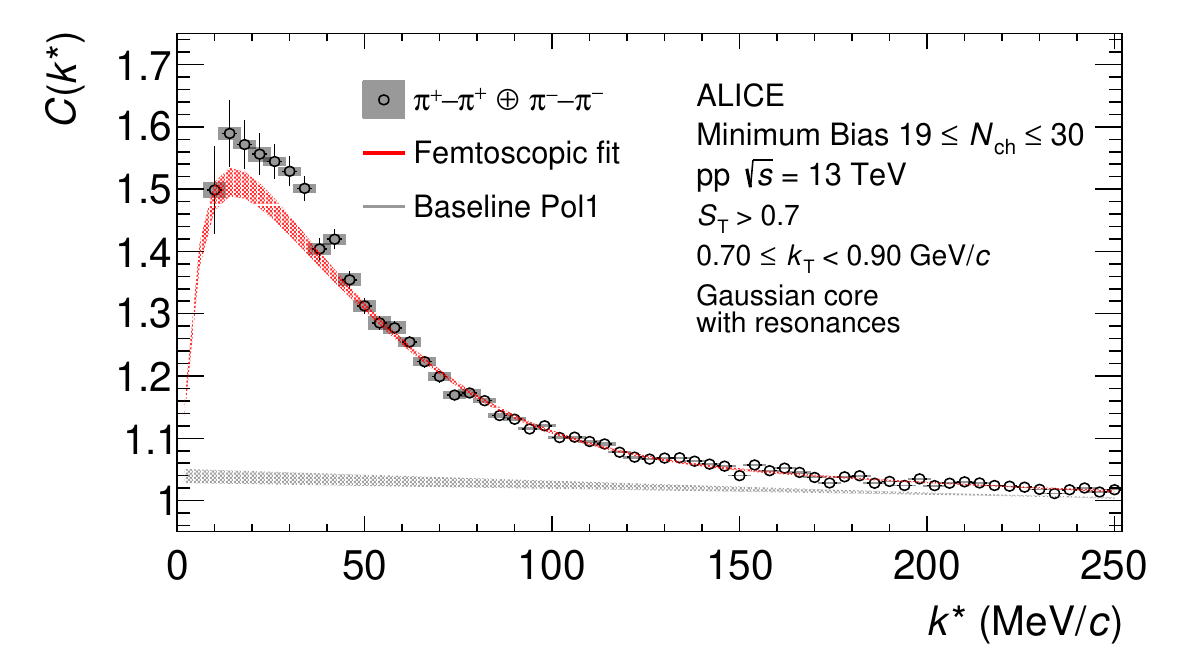}
\includegraphics[width=.47\textwidth]{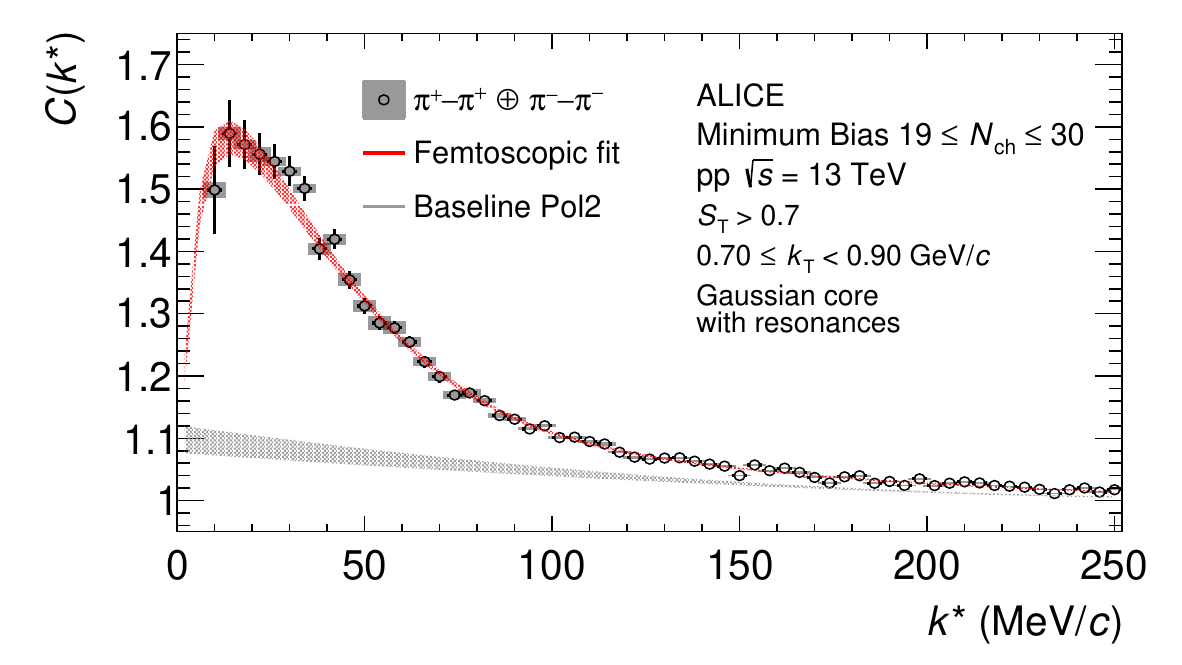}
\includegraphics[width=.47\textwidth]{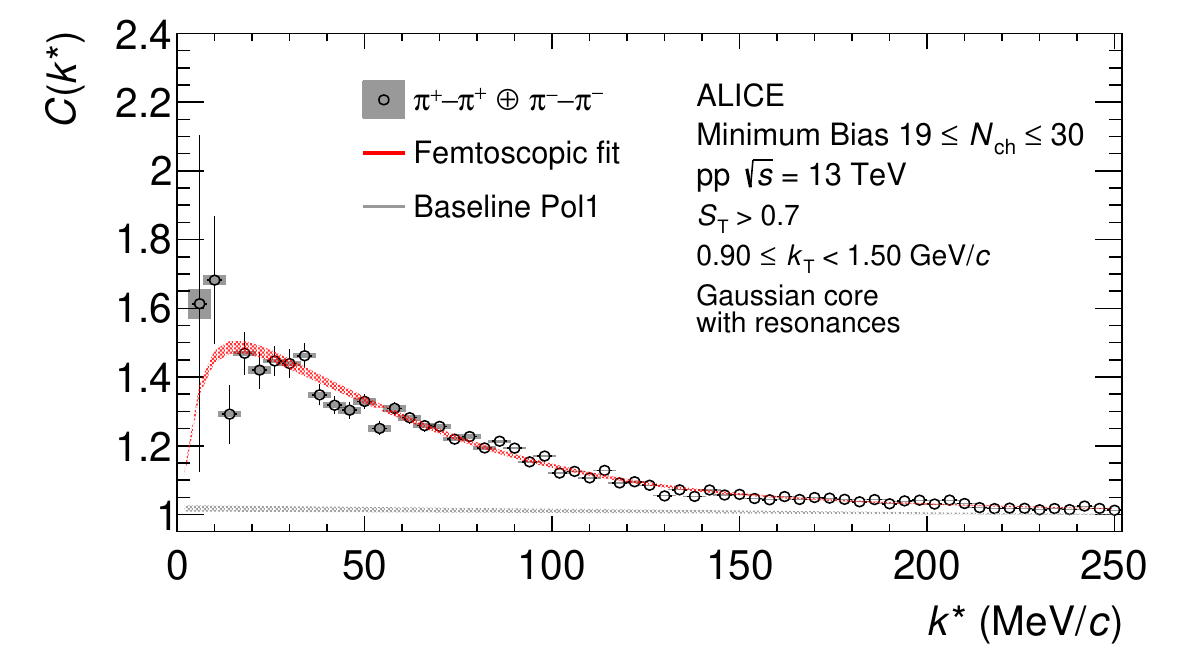}
\includegraphics[width=.47\textwidth]{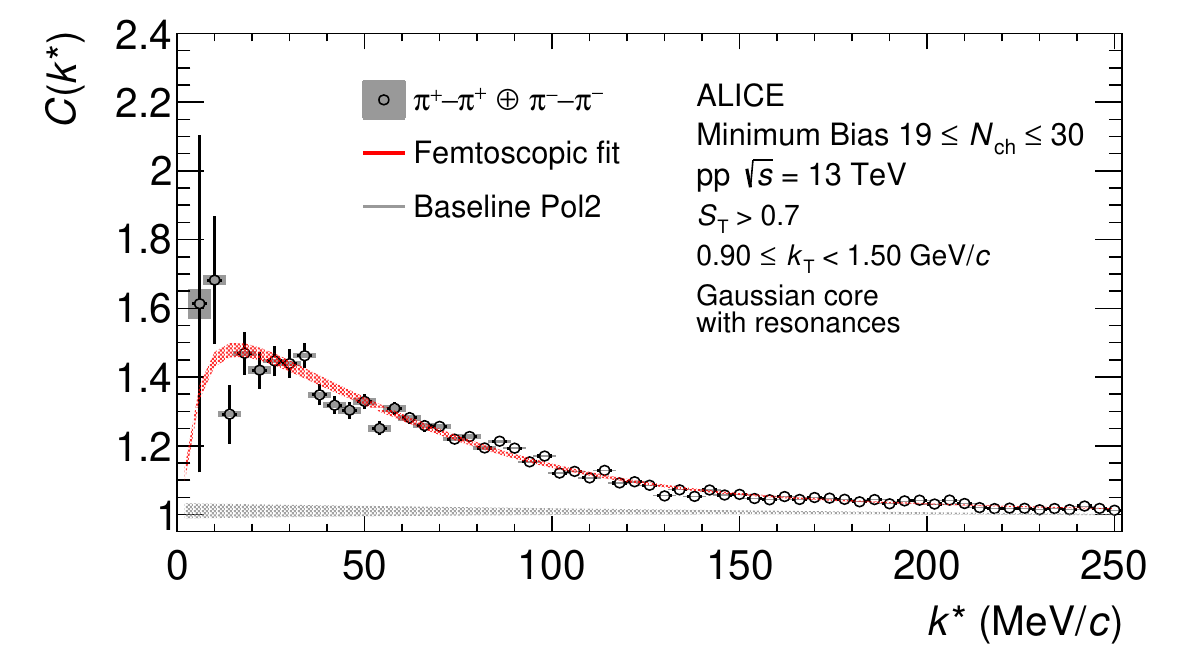}
\caption{Correlation functions of \spipi pairs in bins of \kt for $19\leq~N_{\mathrm{ch}}\leq~30$, the fits are performed using CATS~\cite{CATS} employing the RSM. The left (right) panels show the results assuming a polynomial of first (second) degree as background. The uncertainties of the data points are obtained by varying the selection criteria as described in the text. The uncertainty bands of the fit function are obtained by employing a bootstrap~\cite{Bootstrap} procedure.}
\label{fig:Ck_pipi_MB_mult1}
\end{figure}

\begin{figure}[htbp!]
\centering
\includegraphics[width=.47\textwidth]{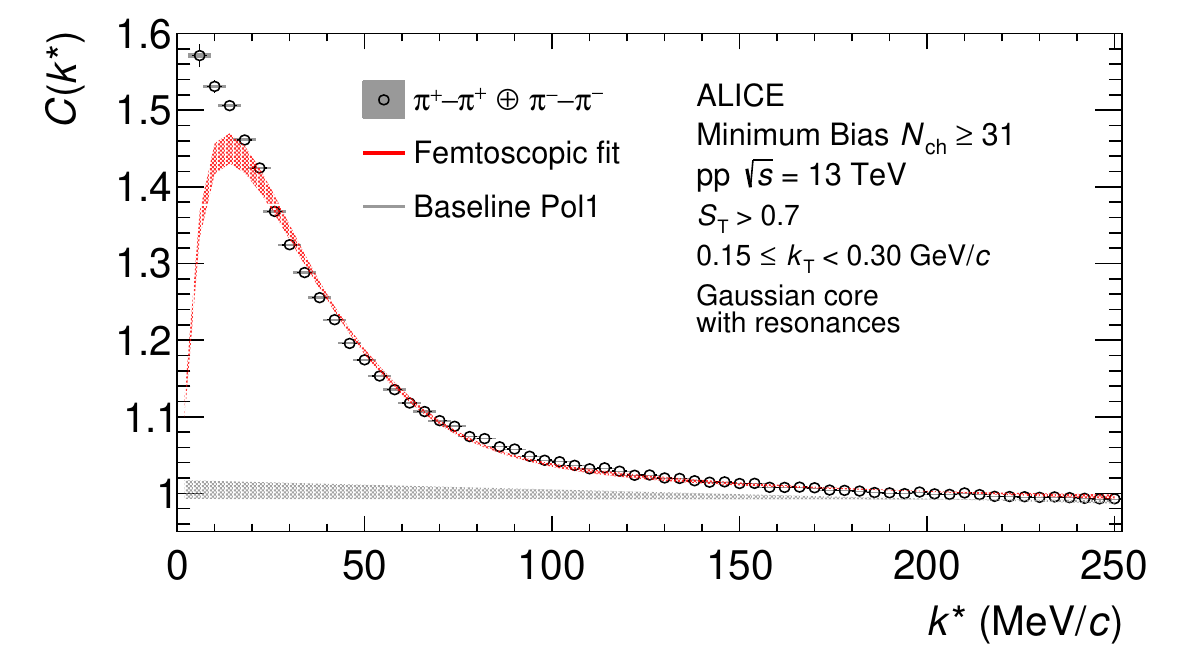}
\includegraphics[width=.47\textwidth]{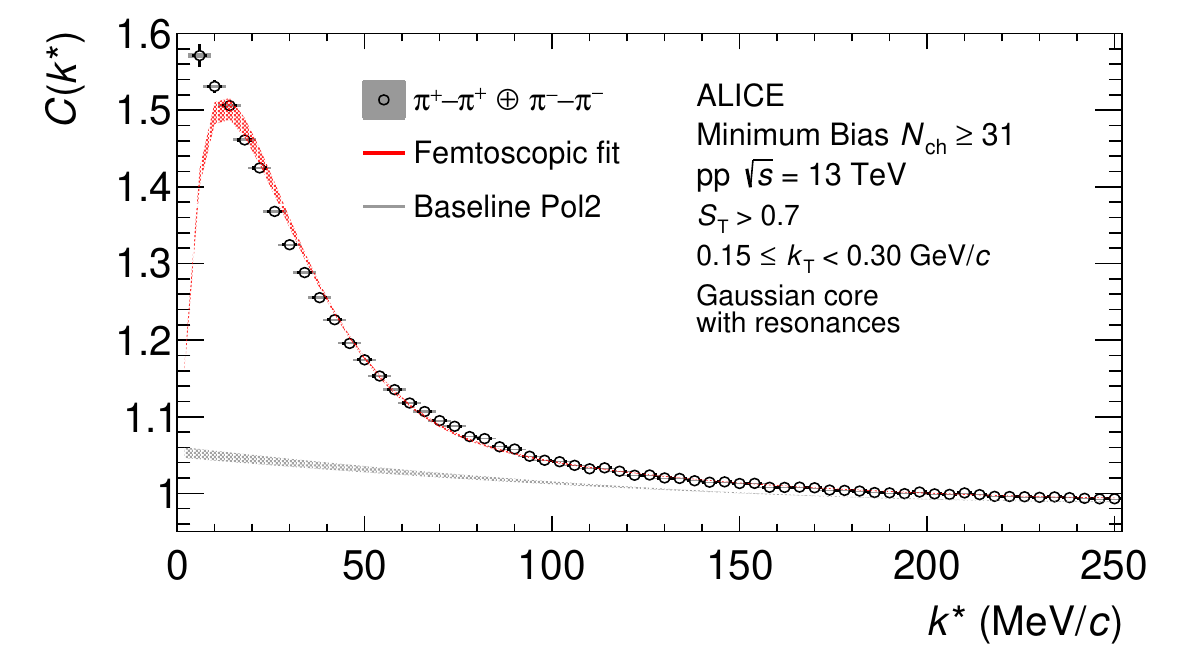}
\includegraphics[width=.47\textwidth]{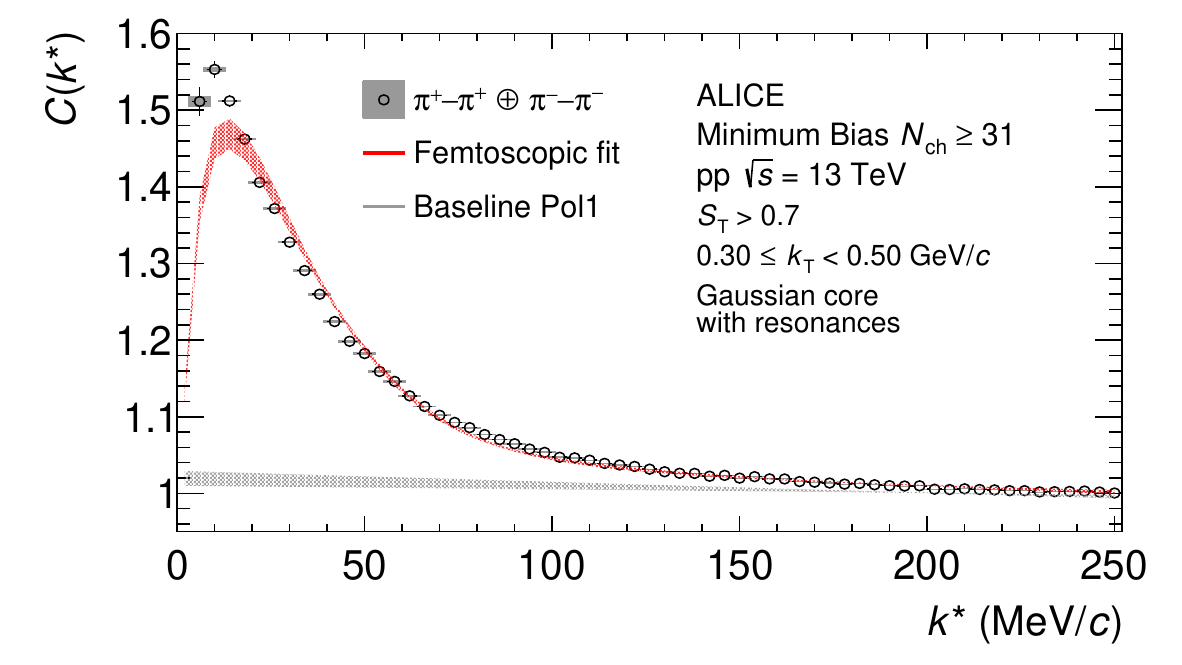}
\includegraphics[width=.47\textwidth]{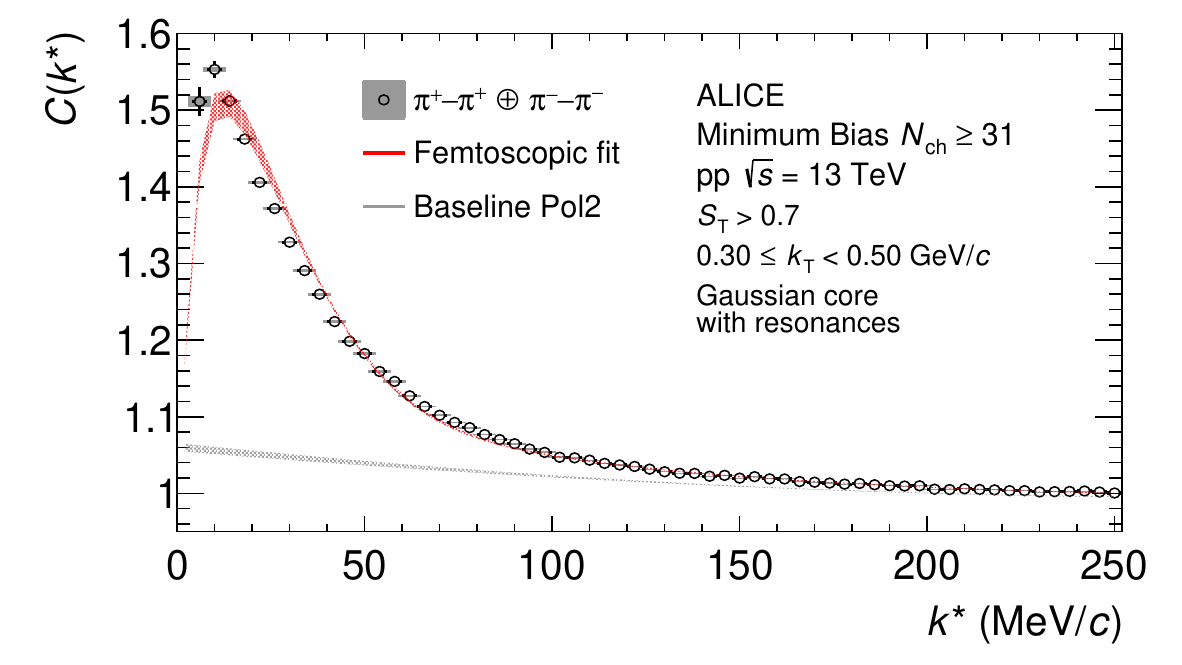}
\includegraphics[width=.47\textwidth]{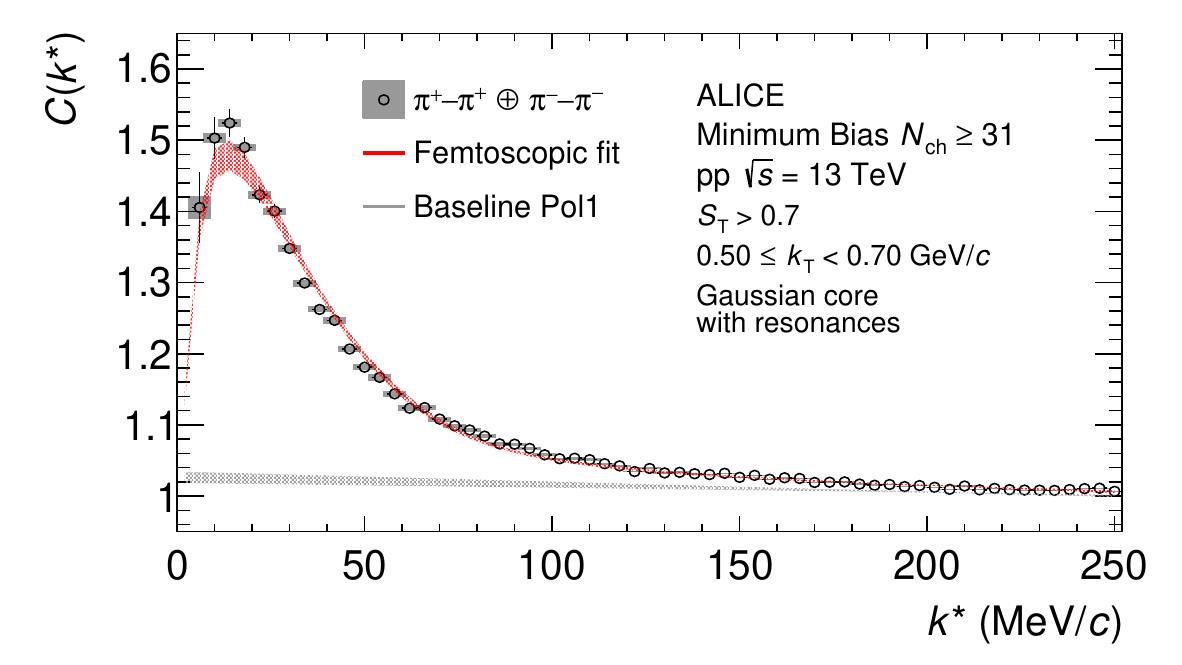}
\includegraphics[width=.47\textwidth]{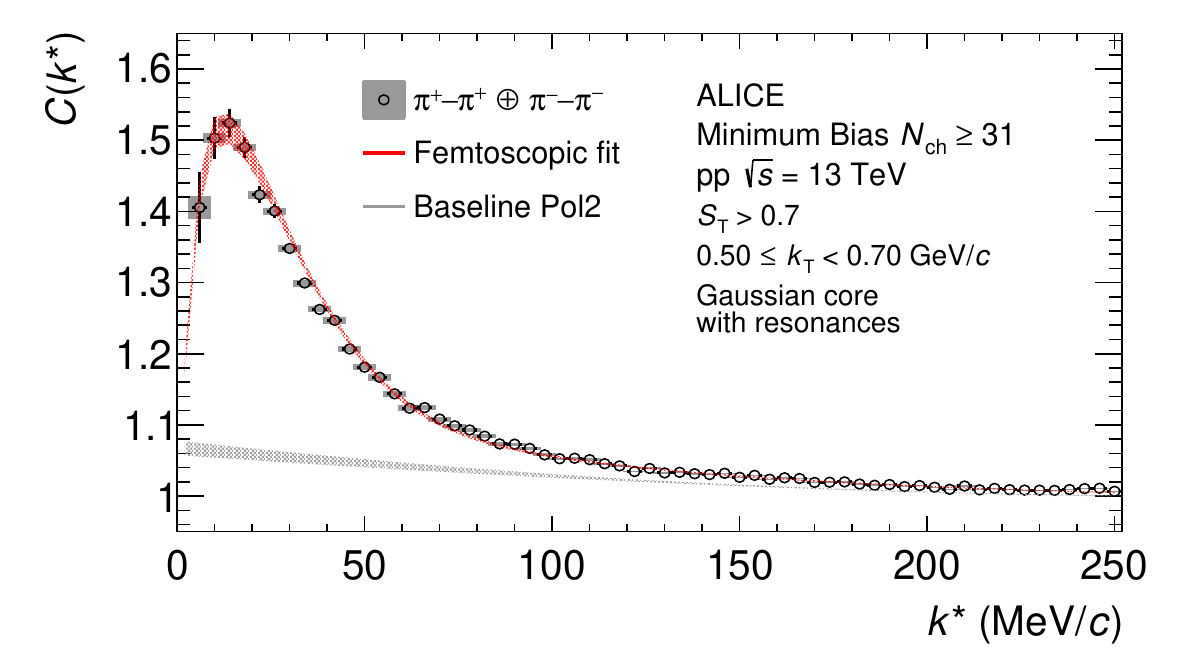}
\includegraphics[width=.47\textwidth]{Figure/E_Results/MB_pipi/Cfs/Plot_13_pol1_new_update.pdf}
\includegraphics[width=.47\textwidth]{Figure/E_Results/MB_pipi/Cfs/Plot_13_pol2_new_update.pdf}
\includegraphics[width=.47\textwidth]{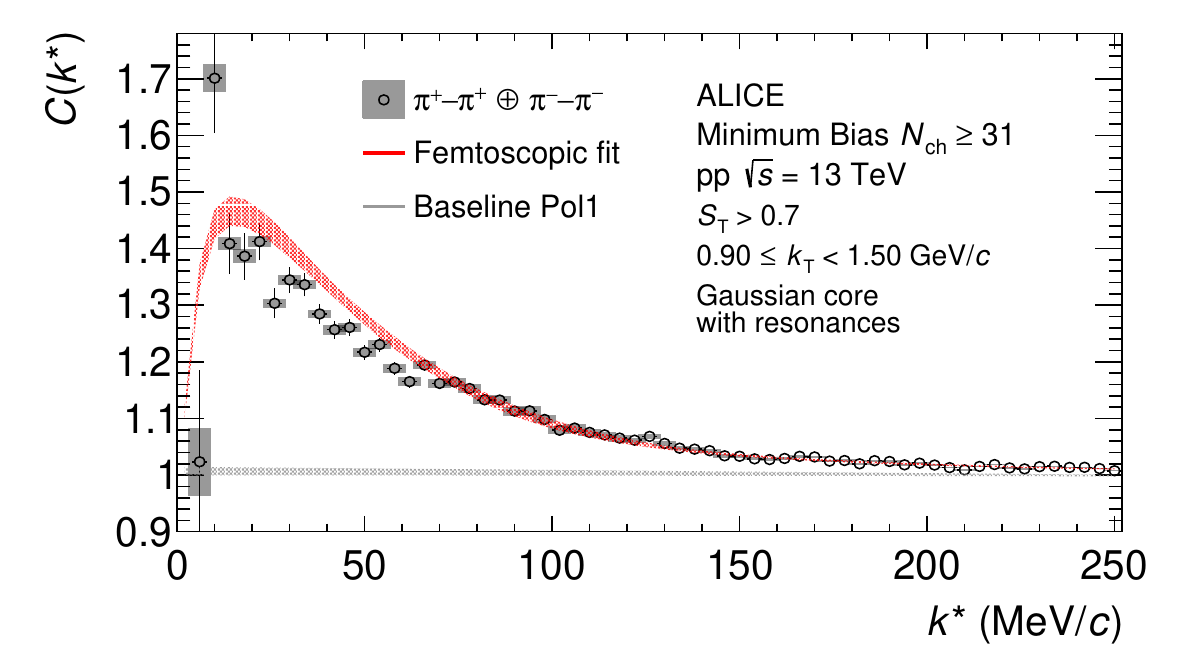}
\includegraphics[width=.47\textwidth]{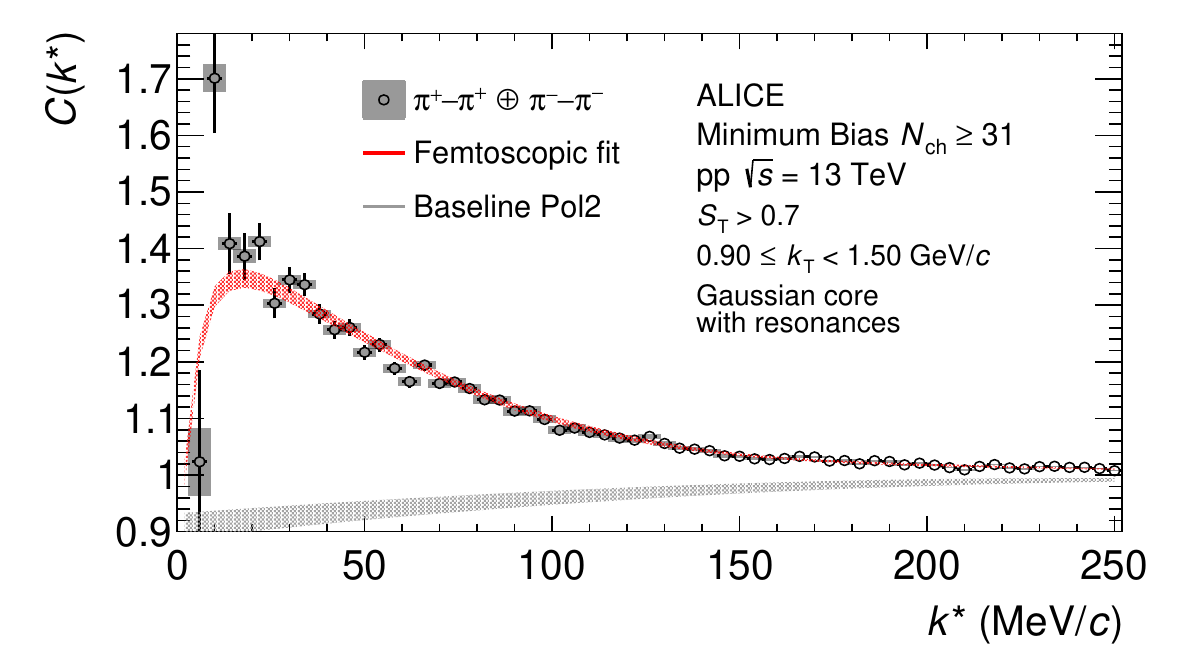}
\caption{Correlation functions of \spipi pairs in bins of \kt for $N_{\mathrm{ch}}\geq~31$, the fits are performed using CATS~\cite{CATS} employing the RSM. The left (right) panels show the results assuming a polynomial of first (second) degree as background. The uncertainties of the data points are obtained by varying the selection criteria as described in the text. The uncertainty bands of the fit function are obtained by employing a bootstrap~\cite{Bootstrap} procedure.}
\label{fig:Ck_pipi_MB_mult2}
\end{figure}

\begin{figure}[htbp!]
\centering
\includegraphics[width=.47\textwidth]{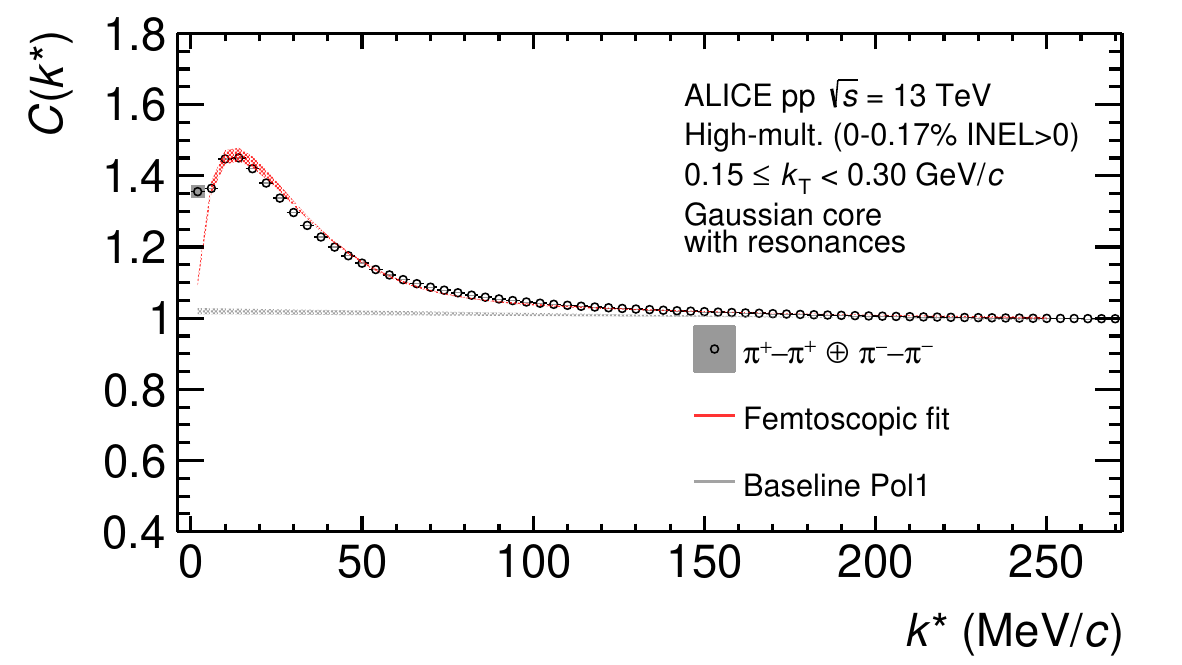}
\includegraphics[width=.47\textwidth]{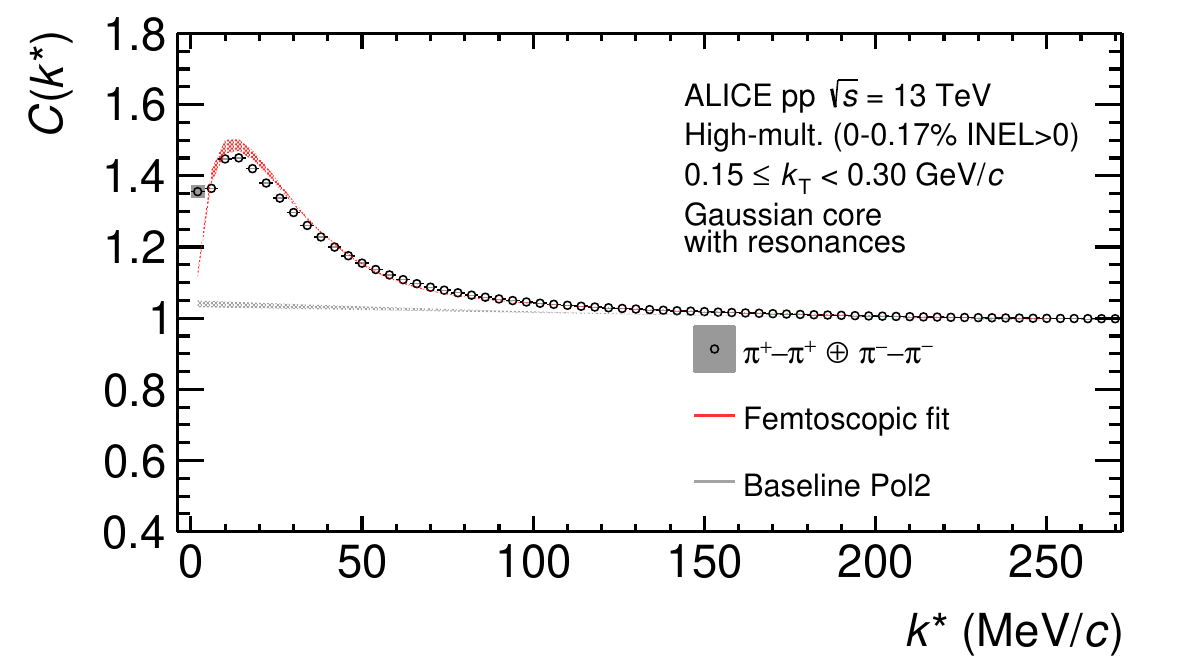}
\includegraphics[width=.47\textwidth]{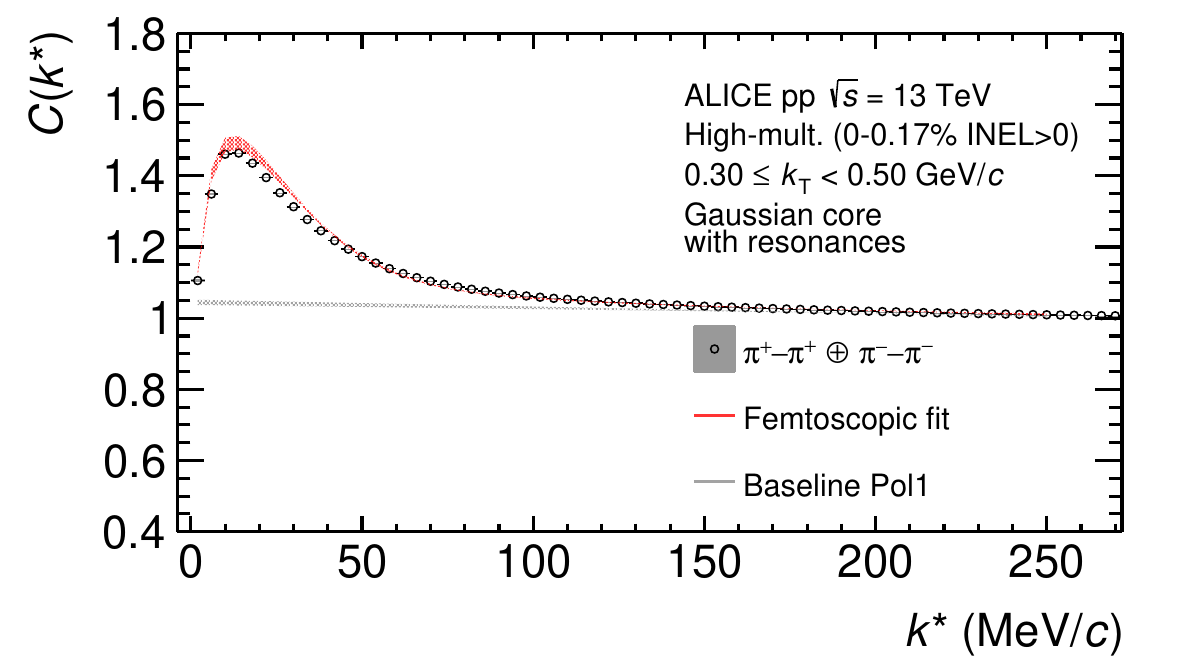}
\includegraphics[width=.47\textwidth]{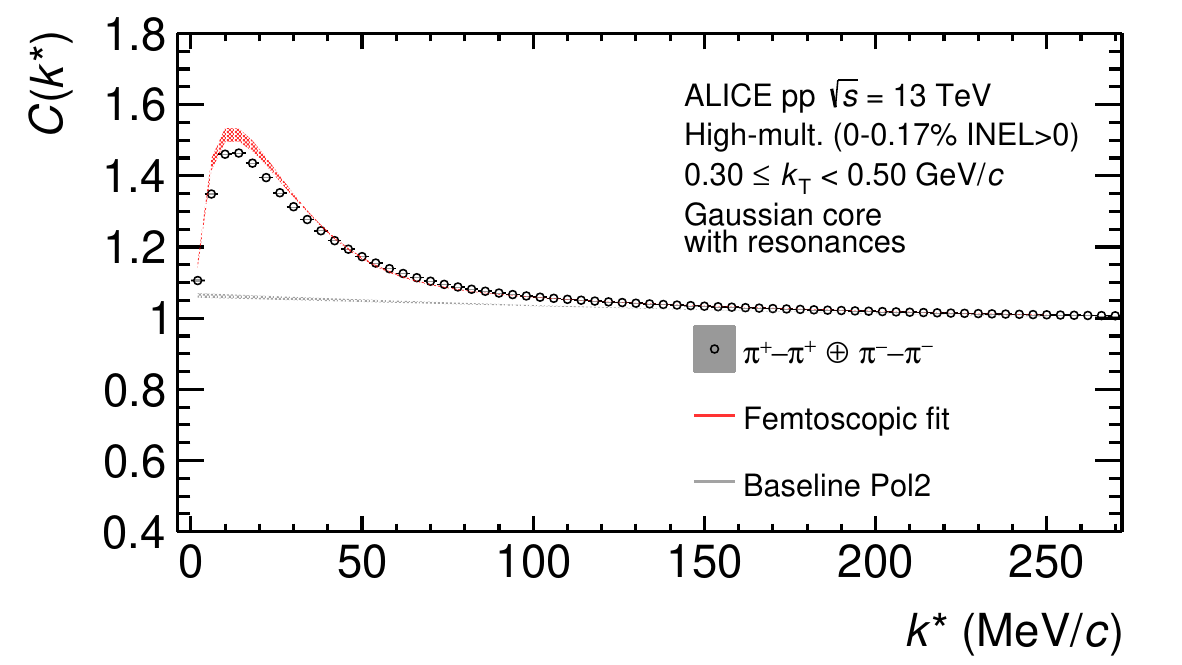}
\includegraphics[width=.47\textwidth]{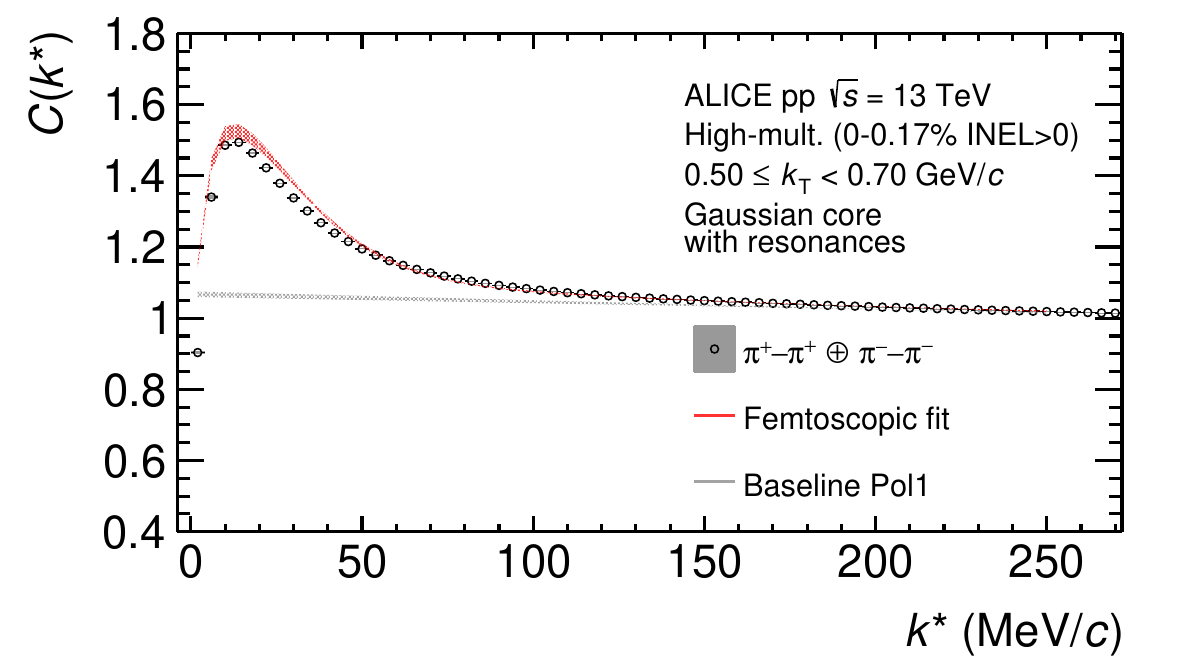}
\includegraphics[width=.47\textwidth]{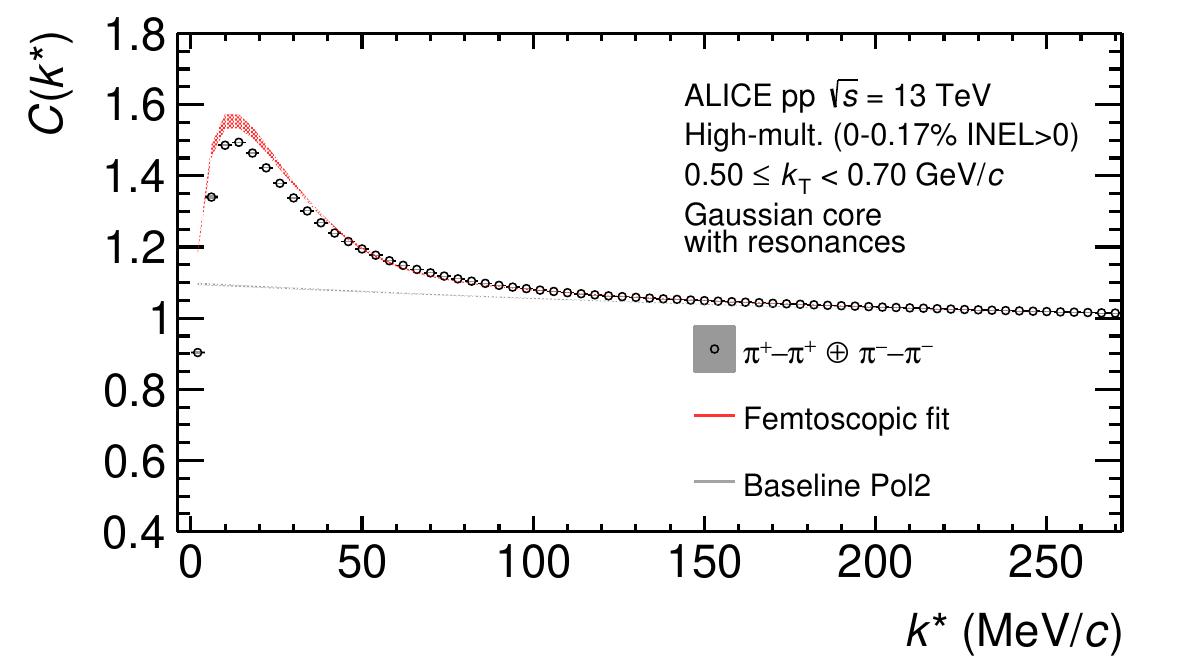}
\includegraphics[width=.47\textwidth]{Figure/E_Results/HM_pipi/Cfs/Default+Syst_ktBin3_Pol1_new_legendUpdate_2.pdf}
\includegraphics[width=.47\textwidth]{Figure/E_Results/HM_pipi/Cfs/Default+Syst_ktBin3_Pol2_new_legendUpdate_2.pdf}
\includegraphics[width=.47\textwidth]{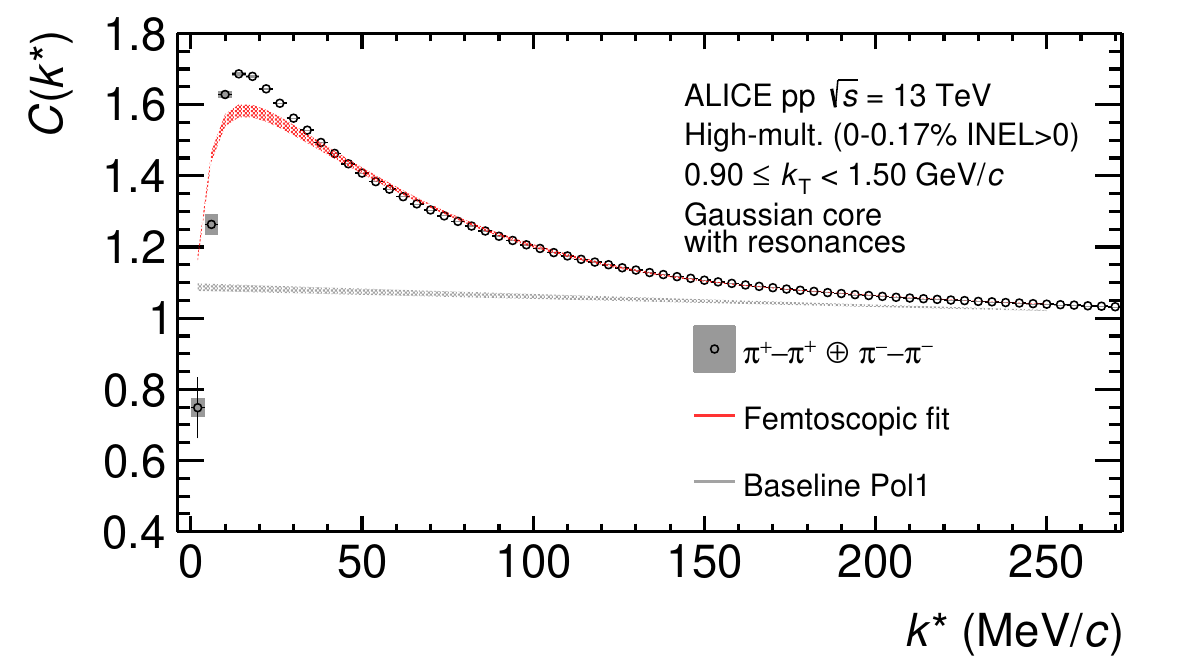}
\includegraphics[width=.47\textwidth]{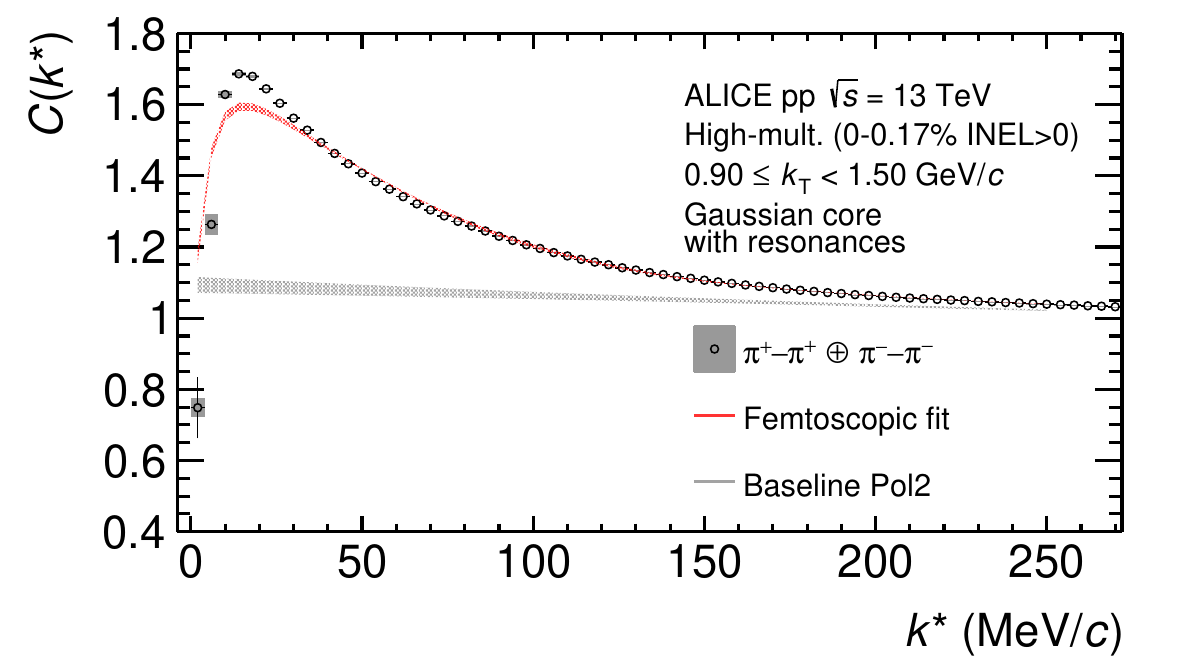}
\caption{Correlation functions of \spipi in bins of \kt for HM collisions, the fits on the left (right) are performed using CATS~\cite{CATS} employing the RSM and assuming a polynomial of first (second) degree as background. The uncertainties of the data points are obtained by varying the selection criteria as described in the text. The uncertainty bands of the fit function are obtained by employing a bootstrap~\cite{Bootstrap} procedure.}
\label{fig:Ck_pipi_HM}
\end{figure}

Correlation functions of same-sign \spipi pairs in bins of \kt are shown in Figs.~\ref{fig:Ck_pipi_MB_mult0},~\ref{fig:Ck_pipi_MB_mult1}, and~\ref{fig:Ck_pipi_MB_mult2} for the three multiplicity classes of MB collisions and in Fig~\ref{fig:Ck_pipi_HM} for the HM dataset.
The systematic uncertainties are represented by boxes.
The fits performed using CATS~\cite{CATS} employing the RSM are also displayed in the figures for the different \kt and multiplicity classes.
The uncertainty bands of the fit function are obtained by employing a bootstrap~\cite{Bootstrap} procedure.
The left (right) panels of the Figures show the results assuming a polynomial of first (second) degree to describe the baseline of the \CF.

%% file: 2023-10-27-Alice_Authorlist_2023-10-27_Opt_C.tex
\begin{flushleft} 
\small

S.~Acharya\,\orcidlink{0000-0002-9213-5329}\,$^{\rm 128}$, 
D.~Adamov\'{a}\,\orcidlink{0000-0002-0504-7428}\,$^{\rm 87}$, 
G.~Aglieri Rinella\,\orcidlink{0000-0002-9611-3696}\,$^{\rm 33}$, 
M.~Agnello\,\orcidlink{0000-0002-0760-5075}\,$^{\rm 30}$, 
N.~Agrawal\,\orcidlink{0000-0003-0348-9836}\,$^{\rm 52}$, 
Z.~Ahammed\,\orcidlink{0000-0001-5241-7412}\,$^{\rm 136}$, 
S.~Ahmad\,\orcidlink{0000-0003-0497-5705}\,$^{\rm 16}$, 
S.U.~Ahn\,\orcidlink{0000-0001-8847-489X}\,$^{\rm 72}$, 
I.~Ahuja\,\orcidlink{0000-0002-4417-1392}\,$^{\rm 38}$, 
A.~Akindinov\,\orcidlink{0000-0002-7388-3022}\,$^{\rm 142}$
% ...

\section*{Affiliation Notes}

$^{\rm I}$ Deceased\\
$^{\rm II}$ Also at: Max-Planck-Institut fur Physik, Munich, Germany\\
$^{\rm III}$ Also at: Italian National Agency for New Technologies, Energy and Sustainable Economic Development (ENEA), Bologna, Italy\\
$^{\rm IV}$ Also at: Dipartimento DET del Politecnico di Torino, Turin, Italy\\
$^{\rm V}$ Also at: Yildiz Technical University, Istanbul, T\"{u}rkiye\\
$^{\rm VI}$ Also at: Department of Applied Physics, Aligarh Muslim University, Aligarh, India\\
$^{\rm VII}$ Also at: Institute of Theoretical Physics, University of Wroclaw, Poland\\
$^{\rm VIII}$ Also at: An institution covered by a cooperation agreement with CERN\\

\section*{Collaboration Institutes}

$^{1}$ A.I. Alikhanyan National Science Laboratory (Yerevan Physics Institute) Foundation, Yerevan, Armenia\\
$^{2}$ AGH University of Krakow, Cracow, Poland\\
$^{3}$ Bogolyubov Institute for Theoretical Physics, National Academy of Sciences of Ukraine, Kiev, Ukraine\\
$^{4}$ Bose Institute, Department of Physics  and Centre for Astroparticle Physics and Space Science (CAPSS), Kolkata, India\\
$^{5}$ California Polytechnic State University, San Luis Obispo, California, United States\\
$^{6}$ Central China Normal University, Wuhan, China\\
$^{7}$ Centro de Aplicaciones Tecnol\'{o}gicas y Desarrollo Nuclear (CEADEN), Havana, Cuba\\
$^{8}$ Centro de Investigaci\'{o}n y de Estudios Avanzados (CINVESTAV), Mexico City and M\'{e}rida, Mexico\\
$^{9}$ Chicago State University, Chicago, Illinois, United States\\
$^{10}$ China Institute of Atomic Energy, Beijing, China\\
$^{11}$ China University of Geosciences, Wuhan, China\\
$^{12}$ Chungbuk National University, Cheongju, Republic of Korea\\
$^{13}$ Comenius University Bratislava, Faculty of Mathematics, Physics and Informatics, Bratislava, Slovak Republic\\
$^{14}$ COMSATS University Islamabad, Islamabad, Pakistan\\
$^{15}$ Creighton University, Omaha, Nebraska, United States\\
$^{16}$ Department of Physics, Aligarh Muslim University, Aligarh, India\\
$^{17}$ Department of Physics, Pusan National University, Pusan, Republic of Korea\\
$^{18}$ Department of Physics, Sejong University, Seoul, Republic of Korea\\
$^{19}$ Department of Physics, University of California, Berkeley, California, United States\\
$^{20}$ Department of Physics, University of Oslo, Oslo, Norway\\
$^{21}$ Department of Physics and Technology, University of Bergen, Bergen, Norway\\
$^{22}$ Dipartimento di Fisica, Universit\`{a} di Pavia, Pavia, Italy\\
$^{23}$ Dipartimento di Fisica dell'Universit\`{a} and Sezione INFN, Cagliari, Italy\\
$^{24}$ Dipartimento di Fisica dell'Universit\`{a} and Sezione INFN, Trieste, Italy\\
$^{25}$ Dipartimento di Fisica dell'Universit\`{a} and Sezione INFN, Turin, Italy\\
$^{26}$ Dipartimento di Fisica e Astronomia dell'Universit\`{a} and Sezione INFN, Bologna, Italy\\
$^{27}$ Dipartimento di Fisica e Astronomia dell'Universit\`{a} and Sezione INFN, Catania, Italy\\
$^{28}$ Dipartimento di Fisica e Astronomia dell'Universit\`{a} and Sezione INFN, Padova, Italy\\
$^{29}$ Dipartimento di Fisica `E.R.~Caianiello' dell'Universit\`{a} and Gruppo Collegato INFN, Salerno, Italy\\
$^{30}$ Dipartimento DISAT del Politecnico and Sezione INFN, Turin, Italy\\
$^{31}$ Dipartimento di Scienze MIFT, Universit\`{a} di Messina, Messina, Italy\\
$^{32}$ Dipartimento Interateneo di Fisica `M.~Merlin' and Sezione INFN, Bari, Italy\\
$^{33}$ European Organization for Nuclear Research (CERN), Geneva, Switzerland\\
$^{34}$ Faculty of Electrical Engineering, Mechanical Engineering and Naval Architecture, University of Split, Split, Croatia\\
$^{35}$ Faculty of Engineering and Science, Western Norway University of Applied Sciences, Bergen, Norway\\
$^{36}$ Faculty of Nuclear Sciences and Physical Engineering, Czech Technical University in Prague, Prague, Czech Republic\\
$^{37}$ Faculty of Physics, Sofia University, Sofia, Bulgaria\\
$^{38}$ Faculty of Science, P.J.~\v{S}af\'{a}rik University, Ko\v{s}ice, Slovak Republic\\
$^{39}$ Frankfurt Institute for Advanced Studies, Johann Wolfgang Goethe-Universit\"{a}t Frankfurt, Frankfurt, Germany\\
$^{40}$ Fudan University, Shanghai, China\\
$^{41}$ Gangneung-Wonju National University, Gangneung, Republic of Korea\\
$^{42}$ Gauhati University, Department of Physics, Guwahati, India\\
$^{43}$ Helmholtz-Institut f\"{u}r Strahlen- und Kernphysik, Rheinische Friedrich-Wilhelms-Universit\"{a}t Bonn, Bonn, Germany\\
$^{44}$ Helsinki Institute of Physics (HIP), Helsinki, Finland\\
$^{45}$ High Energy Physics Group,  Universidad Aut\'{o}noma de Puebla, Puebla, Mexico\\
$^{46}$ Horia Hulubei National Institute of Physics and Nuclear Engineering, Bucharest, Romania\\
$^{47}$ HUN-REN Wigner Research Centre for Physics, Budapest, Hungary\\
$^{48}$ Indian Institute of Technology Bombay (IIT), Mumbai, India\\
$^{49}$ Indian Institute of Technology Indore, Indore, India\\
$^{50}$ INFN, Laboratori Nazionali di Frascati, Frascati, Italy\\
$^{51}$ INFN, Sezione di Bari, Bari, Italy\\
$^{52}$ INFN, Sezione di Bologna, Bologna, Italy\\
$^{53}$ INFN, Sezione di Cagliari, Cagliari, Italy\\
$^{54}$ INFN, Sezione di Catania, Catania, Italy\\
$^{55}$ INFN, Sezione di Padova, Padova, Italy\\
$^{56}$ INFN, Sezione di Pavia, Pavia, Italy\\
$^{57}$ INFN, Sezione di Torino, Turin, Italy\\
$^{58}$ INFN, Sezione di Trieste, Trieste, Italy\\
$^{59}$ Inha University, Incheon, Republic of Korea\\
$^{60}$ Institute for Gravitational and Subatomic Physics (GRASP), Utrecht University/Nikhef, Utrecht, Netherlands\\
$^{61}$ Institute of Experimental Physics, Slovak Academy of Sciences, Ko\v{s}ice, Slovak Republic\\
$^{62}$ Institute of Physics, Homi Bhabha National Institute, Bhubaneswar, India\\
$^{63}$ Institute of Physics of the Czech Academy of Sciences, Prague, Czech Republic\\
$^{64}$ Institute of Space Science (ISS), Bucharest, Romania\\
$^{65}$ Institut f\"{u}r Kernphysik, Johann Wolfgang Goethe-Universit\"{a}t Frankfurt, Frankfurt, Germany\\
$^{66}$ Instituto de Ciencias Nucleares, Universidad Nacional Aut\'{o}noma de M\'{e}xico, Mexico City, Mexico\\
$^{67}$ Instituto de F\'{i}sica, Universidade Federal do Rio Grande do Sul (UFRGS), Porto Alegre, Brazil\\
$^{68}$ Instituto de F\'{\i}sica, Universidad Nacional Aut\'{o}noma de M\'{e}xico, Mexico City, Mexico\\
$^{69}$ iThemba LABS, National Research Foundation, Somerset West, South Africa\\
$^{70}$ Jeonbuk National University, Jeonju, Republic of Korea\\
$^{71}$ Johann-Wolfgang-Goethe Universit\"{a}t Frankfurt Institut f\"{u}r Informatik, Fachbereich Informatik und Mathematik, Frankfurt, Germany\\
$^{72}$ Korea Institute of Science and Technology Information, Daejeon, Republic of Korea\\
$^{73}$ KTO Karatay University, Konya, Turkey\\
$^{74}$ Laboratoire de Physique Subatomique et de Cosmologie, Universit\'{e} Grenoble-Alpes, CNRS-IN2P3, Grenoble, France\\
$^{75}$ Lawrence Berkeley National Laboratory, Berkeley, California, United States\\
$^{76}$ Lund University Department of Physics, Division of Particle Physics, Lund, Sweden\\
$^{77}$ Nagasaki Institute of Applied Science, Nagasaki, Japan\\
$^{78}$ Nara Women{'}s University (NWU), Nara, Japan\\
$^{79}$ National and Kapodistrian University of Athens, School of Science, Department of Physics , Athens, Greece\\
$^{80}$ National Centre for Nuclear Research, Warsaw, Poland\\
$^{81}$ National Institute of Science Education and Research, Homi Bhabha National Institute, Jatni, India\\
$^{82}$ National Nuclear Research Center, Baku, Azerbaijan\\
$^{83}$ National Research and Innovation Agency - BRIN, Jakarta, Indonesia\\
$^{84}$ Niels Bohr Institute, University of Copenhagen, Copenhagen, Denmark\\
$^{85}$ Nikhef, National institute for subatomic physics, Amsterdam, Netherlands\\
$^{86}$ Nuclear Physics Group, STFC Daresbury Laboratory, Daresbury, United Kingdom\\
$^{87}$ Nuclear Physics Institute of the Czech Academy of Sciences, Husinec-\v{R}e\v{z}, Czech Republic\\
$^{88}$ Oak Ridge National Laboratory, Oak Ridge, Tennessee, United States\\
$^{89}$ Ohio State University, Columbus, Ohio, United States\\
$^{90}$ Physics department, Faculty of science, University of Zagreb, Zagreb, Croatia\\
$^{91}$ Physics Department, Panjab University, Chandigarh, India\\
$^{92}$ Physics Department, University of Jammu, Jammu, India\\
$^{93}$ Physics Program and International Institute for Sustainability with Knotted Chiral Meta Matter (SKCM2), Hiroshima University, Hiroshima, Japan\\
$^{94}$ Physikalisches Institut, Eberhard-Karls-Universit\"{a}t T\"{u}bingen, T\"{u}bingen, Germany\\
$^{95}$ Physikalisches Institut, Ruprecht-Karls-Universit\"{a}t Heidelberg, Heidelberg, Germany\\
$^{96}$ Physik Department, Technische Universit\"{a}t M\"{u}nchen, Munich, Germany\\
$^{97}$ Politecnico di Bari and Sezione INFN, Bari, Italy\\
$^{98}$ Research Division and ExtreMe Matter Institute EMMI, GSI Helmholtzzentrum f\"ur Schwerionenforschung GmbH, Darmstadt, Germany\\
$^{99}$ Saga University, Saga, Japan\\
$^{100}$ Saha Institute of Nuclear Physics, Homi Bhabha National Institute, Kolkata, India\\
$^{101}$ School of Physics and Astronomy, University of Birmingham, Birmingham, United Kingdom\\
$^{102}$ Secci\'{o}n F\'{\i}sica, Departamento de Ciencias, Pontificia Universidad Cat\'{o}lica del Per\'{u}, Lima, Peru\\
$^{103}$ Stefan Meyer Institut f\"{u}r Subatomare Physik (SMI), Vienna, Austria\\
$^{104}$ SUBATECH, IMT Atlantique, Nantes Universit\'{e}, CNRS-IN2P3, Nantes, France\\
$^{105}$ Sungkyunkwan University, Suwon City, Republic of Korea\\
$^{106}$ Suranaree University of Technology, Nakhon Ratchasima, Thailand\\
$^{107}$ Technical University of Ko\v{s}ice, Ko\v{s}ice, Slovak Republic\\
$^{108}$ The Henryk Niewodniczanski Institute of Nuclear Physics, Polish Academy of Sciences, Cracow, Poland\\
$^{109}$ The University of Texas at Austin, Austin, Texas, United States\\
$^{110}$ Universidad Aut\'{o}noma de Sinaloa, Culiac\'{a}n, Mexico\\
$^{111}$ Universidade de S\~{a}o Paulo (USP), S\~{a}o Paulo, Brazil\\
$^{112}$ Universidade Estadual de Campinas (UNICAMP), Campinas, Brazil\\
$^{113}$ Universidade Federal do ABC, Santo Andre, Brazil\\
$^{114}$ Universitatea Nationala de Stiinta si Tehnologie Politehnica Bucuresti, Bucharest, Romania\\
$^{115}$ University of Cape Town, Cape Town, South Africa\\
$^{116}$ University of Derby, Derby, United Kingdom\\
$^{117}$ University of Houston, Houston, Texas, United States\\
$^{118}$ University of Jyv\"{a}skyl\"{a}, Jyv\"{a}skyl\"{a}, Finland\\
$^{119}$ University of Kansas, Lawrence, Kansas, United States\\
$^{120}$ University of Liverpool, Liverpool, United Kingdom\\
$^{121}$ University of Science and Technology of China, Hefei, China\\
$^{122}$ University of South-Eastern Norway, Kongsberg, Norway\\
$^{123}$ University of Tennessee, Knoxville, Tennessee, United States\\
$^{124}$ University of the Witwatersrand, Johannesburg, South Africa\\
$^{125}$ University of Tokyo, Tokyo, Japan\\
$^{126}$ University of Tsukuba, Tsukuba, Japan\\
$^{127}$ Universit\"{a}t M\"{u}nster, Institut f\"{u}r Kernphysik, M\"{u}nster, Germany\\
$^{128}$ Universit\'{e} Clermont Auvergne, CNRS/IN2P3, LPC, Clermont-Ferrand, France\\
$^{129}$ Universit\'{e} de Lyon, CNRS/IN2P3, Institut de Physique des 2 Infinis de Lyon, Lyon, France\\
$^{130}$ Universit\'{e} de Strasbourg, CNRS, IPHC UMR 7178, F-67000 Strasbourg, France, Strasbourg, France\\
$^{131}$ Universit\'{e} Paris-Saclay, Centre d'Etudes de Saclay (CEA), IRFU, D\'{e}partment de Physique Nucl\'{e}aire (DPhN), Saclay, France\\
$^{132}$ Universit\'{e}  Paris-Saclay, CNRS/IN2P3, IJCLab, Orsay, France\\
$^{133}$ Universit\`{a} degli Studi di Foggia, Foggia, Italy\\
$^{134}$ Universit\`{a} del Piemonte Orientale, Vercelli, Italy\\
$^{135}$ Universit\`{a} di Brescia, Brescia, Italy\\
$^{136}$ Variable Energy Cyclotron Centre, Homi Bhabha National Institute, Kolkata, India\\
$^{137}$ Warsaw University of Technology, Warsaw, Poland\\
$^{138}$ Wayne State University, Detroit, Michigan, United States\\
$^{139}$ Yale University, New Haven, Connecticut, United States\\
$^{140}$ Yonsei University, Seoul, Republic of Korea\\
$^{141}$  Zentrum  f\"{u}r Technologie und Transfer (ZTT), Worms, Germany\\
$^{142}$ Affiliated with an institute covered by a cooperation agreement with CERN\\
$^{143}$ Affiliated with an international laboratory covered by a cooperation agreement with CERN.\\

\end{flushleft} 